\documentclass[11pt,english]{article}
\usepackage[LGR,LGR,LGR,LGR,LGR,LGR,LGR,T1]{fontenc}
\usepackage[latin9]{inputenc}
\usepackage{geometry}
\usepackage{array}
\usepackage{url}
\usepackage{multirow}
\usepackage{amsmath}
\usepackage{amssymb}
\usepackage{graphicx}
\usepackage{setspace}
\makeatletter

\DeclareRobustCommand{\greektext}{%
  \fontencoding{LGR}\selectfont\def\encodingdefault{LGR}}
\DeclareRobustCommand{\textgreek}[1]{\leavevmode{\greektext #1}}
\ProvideTextCommand{\~}{LGR}[1]{\char126#1}

\providecommand{\tabularnewline}{\\}

\@ifundefined{date}{}{\date{}}
\usepackage{amsfonts}
\usepackage{colortbl}
\definecolor{dgray}{gray}{0.5}

\usepackage{etoolbox}
\usepackage{babel}

\usepackage[font={small}]{caption}

\usepackage{babel}

\usepackage{titling}

\usepackage{algorithm,algpseudocode}

\newcommand*{\algrule}[1][\algorithmicindent]{%
  \makebox[#1][l]{%
    \hspace*{.2em}% <------------- This is where the rule starts from
    \vrule height .75\baselineskip depth .25\baselineskip
  }
}

\newcount\ALG@printindent@tempcnta
\def\ALG@printindent{%
    \ifnum \theALG@nested>0% is there anything to print
    \ifx\ALG@text\ALG@x@notext% is this an end group without any text?
    \else
    \unskip
    \ALG@printindent@tempcnta=1
    \loop
    \algrule[\csname ALG@ind@\the\ALG@printindent@tempcnta\endcsname]%
    \advance \ALG@printindent@tempcnta 1
    \ifnum \ALG@printindent@tempcnta<\numexpr\theALG@nested+1\relax
    \repeat
    \fi
    \fi
}
\patchcmd{\ALG@doentity}{\noindent\hskip\ALG@tlm}{\ALG@printindent}{}{\errmessage{failed to patch}}
\patchcmd{\ALG@doentity}{\item[]\nointerlineskip}{}{}{} % no spurious vertical space
\algnewcommand\algorithmicinput{\textbf{Input:}}
\algnewcommand\INPUT{\item[\algorithmicinput]}

\algnewcommand{\Initialize}[1]{%
  \State \textbf{Initialize:}
  \Statex \hspace*{\algorithmicindent}\parbox[t]{.8\linewidth}{\raggedright #1}
}

\algnewcommand\algorithmicoutput{\textbf{Output:}}
\algnewcommand\OUTPUT{\item[\algorithmicoutput]}

\algnewcommand{\Output}[1]{%
  \State \textbf{Output:}
  \Statex \hspace*{\algorithmicindent}\parbox[t]{.8\linewidth}{\raggedright #1}
}

\usepackage{tikz}

\usepackage [numbers]{natbib}

\@ifundefined{showcaptionsetup}{}{%
 \PassOptionsToPackage{caption=false}{subfig}}
\usepackage{subfig}
\makeatother

\usepackage{babel}
\begin{document}
\vspace{-2.0cm}

\title{The Impact of Estimation: A New Method for Clustering and Trajectory
Estimation in Patient Flow Modeling}

\maketitle
\vspace{-4.0cm}

\noindent\begin{minipage}[t]{1\columnwidth}%
\begin{singlespace}
\begin{center}
{\footnotesize{}Chitta Ranjan\textsuperscript{\dag {*}}, Kamran Paynabar\textsuperscript{\dag{}},
Jonathan E. Helm\textsuperscript{\ddag{}} and Julian Pan\textsuperscript{\textasciitilde{}}}
\par\end{center}{\footnotesize \par}
\begin{center}
{\footnotesize{}\textsuperscript{{*}}Email: \url{nk.chitta.ranjan@gatech.edu}}
\par\end{center}{\footnotesize \par}
\begin{center}
{\footnotesize{}\textsuperscript{\dag{} }H. Milton Stewart School
of Industrial and Systems Engineering, Georgia Institute of Technology,
Atlanta, GA}
\par\end{center}{\footnotesize \par}
\begin{center}
{\footnotesize{}\textsuperscript{\ddag{} }Kelley School of Business,
Indiana University, Bloomington, IN}
\par\end{center}{\footnotesize \par}
\begin{center}
{\footnotesize{}\textsuperscript{\textasciitilde{}}Lean Care Solutions
Corporation Pte. Ltd. 28 Ayer Rajah Crescent \#03-01, Singapore 139959}
\par\end{center}{\footnotesize \par}
\end{singlespace}
\medskip{}

\end{minipage}
\begin{abstract}
{\footnotesize{}{}{}{}{}{}The ability to accurately forecast
and control inpatient census, and thereby workloads, is a critical
and longstanding problem in hospital management. The majority of current
literature focuses on optimal scheduling of inpatients, but largely
ignores the process of accurate estimation of the trajectory of patients
throughout the treatment and recovery process. The result is that
current scheduling models are optimizing based on inaccurate input
data. We developed a Clustering and Scheduling Integrated (CSI) approach
to capture patient flows through a network of hospital services. CSI
functions by clustering patients into groups based on similarity of
trajectory using a novel Semi-Markov model (SMM)-based clustering
scheme, as opposed to clustering by patient attributes 
as in previous literature. Our methodology is validated
by simulation and then applied to real patient data from a partner
hospital where we demonstrate that it outperforms a suite of well-established 
clustering methods. Further, we
demonstrate that extant optimization methods achieve significantly
better results on key hospital performance measures under CSI, compared
with traditional estimation approaches, increasing elective admissions
by 97\% and utilization by 22\% compared to 30\% and 8\% using traditional
estimation techniques. From a theoretical standpoint, the SMM-clustering
is a novel approach applicable to any temporal-spatial stochastic
data that is prevalent in many industries and application areas.}{\footnotesize \par}
\end{abstract}
{\footnotesize{}{}{}{}{}{}{}}\textbf{\footnotesize{}{}{}{}{}{}Keywords:}{\footnotesize{}{}{}{}{}{}
Clustering, EM algorithm, semi-Markov mixture model, patient flow
estimation, stochastic location models.}{\footnotesize \par}

\vspace{-0.2in}

\section{Introduction}

\vspace{-0.2in}

The mismatch between demand for and supply of medical services caused
by high hospital census variability has challenged hospital managers
for decades. High census variability is a common problem in hospitals
and healthcare centers around the world. This problem leads to poor
quality of care, blocking in hospital wards, increase in inpatient
length of stay, and ultimately causes significant increase in cost
for both patient and hospital (Helm and Van Oyen 2015). Aiken et al.
(2002) studied the effect of overloaded nursing staff induced by census
variability and showed its effect on mortality rate, nurse burnout
and job dissatisfaction. A common approach to managing census variability
in practice involves hospitals procuring excess resources including
material, staff, and equipment, leading to frequent instances of under-utilization
for very expensive resources (Griffin et al. 2012). A better approach
is to optimize the utilization of available hospital resources based
on patient census estimations. This long-standing problem has been
termed the \emph{Hospital Admission Scheduling and Control} (HASC)
problem, which can be decomposed into two main steps: \emph{census
modeling} (CM) and \emph{resource scheduling} (RS). CM estimates distributional
information (typically mean and variance) on patient census at the
ward level, which is used as an input to the RS to find the optimal
resource allocation plans and schedules for elective inpatient admissions.

A significant body of work addresses the RS through a variety of optimization
approaches, however research on effective census models that integrate
with RS is less developed. In this paper, we develop a CM method that
integrates well with existing RS methods to solve the HASC. We further
demonstrate the importance of the CM component with respect to the
outcomes of the RS optimization; a factor that has, to our knowledge,
been unaddressed in the current literature. Namely, we show, through computational 
experiments and a case study based on data from our industry partner,  
that the CM method 
typically employed in RS optimization papers leads to markedly inferior optimization
results. To conclude this section, we give
a short description of the current state of the hospital census forecasting
and optimization industry from the experiences of our industry co-author
and CEO of a healthcare analytics company. Then we discuss challenges
posed by the gaps in CM theory that represent a major hurdle for this
burgeoning industry and discuss how our approach seeks to bridge those
gaps.

\subsection{Real-world Challenges in the Hospital Census Forecasting Industry}

\vspace{-0.2in}

Predicting future hospital census levels is a key challenge in the
Hospital Admission Scheduling and Control (HASC) problem. Without
accurate forecasting mechanisms, controlling the variability in hospital
census becomes difficult and creates a major barrier to low cost, high quality inpatient
care. These consequences of inadequate forecasting are drawn from
real-world experience, where our co-author has worked with clients
and collaborators globally - Asia, Europe and North America. All the
hospitals he has worked with experience significant mid-week congestion
and high levels of blocking.

Current methodologies used in hospitals are ineffective to solving
the HASC problem. Almost all the hospitals have lean teams focused
on process improvement and some of the bigger hospitals have small
analytics teams that use rudimentary models which are ineffective
at implementing changes made to solve the HASC problem. All the work
done at the hospital level are reactive models (predicting census
levels using historical census means, and applying control by canceling
surgeries the day before) versus proactive models (implementing control
measures in advance). Recently, some hospitals have been attempting
to shift to proactive measures. This has typically involved increasing
capacity and lowering utilization, which is cost prohibitive in the
long-run. The real solution is to improve the forecasting technology.
The methods outlined in this paper have proven to be effective on
a conceptual level with results shared in the later sections.

Our collaborator, company XYZ (the real name of the company is currently
disguised for the review process), is one of the first to provide
a patient level forecasting tool; i.e. predicting individual flows
and trajectories of each type of patient entering the hospital. A
patient level forecasting and control tool is imperative for hospitals
to effectively solve the HASC problem. While forecasting is the backbone
to the solution, XYZ also provides the ability for hospitals to create
what-if scenarios by modifying admission plans and schedules and to
use optimization techniques to customize a dynamic admission plan
to minimize blocking and surgical cancellations. This type of analysis
and decision support is only possible through patient-level forecasting, 
as it requires understanding how patient-by-patient modifications
to the admission schedule impact hospital census and blocking. This
is precisely the type of forecasting that we propose in this paper.
In fact, workload forecasting is not only useful for bed planning
purposes, but is key to allocating resources to the various functions 
of the hospital. Most notably, workforce planning for front and back
end staff accounts for over 50\% of hospital costs. Based on the
feedback received from XYZ clients, properly allocating staffing reduces
various costs, like overtime, and improves staff satisfaction. Overall,
it is one key in keeping hospitals profitable and delivering top quality
healthcare. After discussing the various needs of the hospitals, it
is clear that the key issues in patient flow management, staffing,
and scheduling all rely on the critical role of forecasting flexibility
and accuracy.

One ongoing challenge for XYZ is the issue of defining Patient Types
(PTypes) and estimating their probabilistic trajectories over the
course of their hospital stay, both of which have a major affect on
forecast accuracy. From a computational standpoint, it requires clustering
patients into groups, where each group represents one type of patient.
Currently, XYZ employs various forms of regressions to determine factors
to group similar patients together into clusters based on patient
characteristics. Many assumptions must be made to fit data into logical
PTypes that are scalable and yet give enough information to statistically
differentiate patients and enable accurate forecasting. This includes
applying numerous heuristics and unfortunately, sacrificing the accuracy
of the forecast. At XYZ, this process is currently done manually for
each hospital, often requiring weeks to months of effort to properly
tailor the PTypes for accurate forecasting. These issues of scalability,
repeatability, and demonstrated statistical accuracy represent one
of the major hurdles for XYZ and other participants in the patient-level
forecasting space. The methods presented in this paper help solve
a key problem in parameterizing models for each hospital. Specifically,
by clustering patients based on trajectory (rather than extrinsic
characteristics as in current practice) this paper significantly improves
upon the currently time consuming and heuristic step of assigning
PTypes. Our approach is shown to be scalable, statistically rigorous,
accurate, and repeatable. This eliminates the time consuming, gestalt
guess work inherent in current practice and has proven to significantly
increase forecast accuracy in addition to improving the results from
current decision support methods for admission scheduling.

\subsection{Failures of Traditional CM Methods.}

\vspace{-0.2in}

\begin{figure}
\begin{centering}
\includegraphics[scale=0.6]{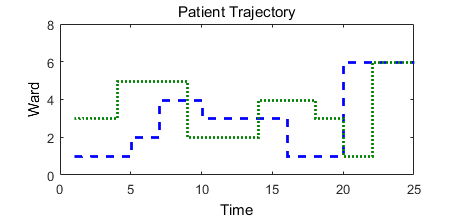} 
\par\end{centering}
\caption{Example of sample path outcomes of a stochastic location process for
patient flow. The x-axis shows the time after admission, while the
y-axis denotes the ward the patient is in at time $t$; each step
is a change of ward for the patient. \label{fig:Example-of-patient}}
\end{figure}

As noted in Fetter et al. (1980) and Helm and Van Oyen (2015), an
appropriate HASC model should have three characteristics: \emph{scalable}
to hospitals of any size, consider \emph{ward interactions}, and account
for \emph{patient heterogeneity}. Most work in the RS step assumes
that patient types are given and uses simple methods for estimating
patient trajectories, then employs analytical techniques to capture
key hospital metrics in an optimization model. A patient trajectory
is characterized by the transitions between wards in a hospital and
patient Length of Stay (LOS) in each ward, and can be expressed as
a stochastic function called a location process that maps time to
a set of locations \textemdash{} see Fig.~\ref{fig:Example-of-patient}
for an example of two sample path outcomes of a location process.

While the optimization methods are generalizable, the previous approaches
to CM for RS optimization lack scalability and are not well suited
for capturing patient heterogeneity. These approaches also suffer from the limitation 
of not properly capturing ward interactions, which is shown to be important in Sec. 
\ref{subsec:Evaluating-the-Impact} and \ref{sec:Case-study-on}, where we compare our method with traditional methods that fail to properly account 
for ward interactions.
In this paper we address these issues by developing
new methods for clustering patient location processes based on historical
patient flow data.

As an example of how clustering impacts scalability and patient heterogeneity,
consider the following. Many traditional approaches to HASC cluster
patients by their diagnosis related group (DRG) or admitting service.
However, in working with a large hospital such as our partner hospital,
there can be close to one hundred such patient types with quite a
few of them being very rare. With such a large number of patient types
we have found that there is insufficient data to properly estimate
patient trajectories even with two or more years of historical data.
When further including other important factors such as gender and
age, which have been found to be important in determining a patient's
trajectory, data scarcity becomes an even larger problem. Current
solutions include combining different patient types that are deemed
\textquotedbl{}similar\textquotedbl{} in order to have sufficient
historical data for trajectory estimation. This is a \emph{clustering}
problem. Deciding how to combine patient types, however, is a non-trivial
effort considering the entire location process (time and location)
must be compared to ensure an accurate pairing of two patient types.
For example, two patient types may have the same average length of
stay (LOS) in the hospital but visit different wards. Another example
is if two patients visit similar locations with similar mean LOS,
but one has a skewed LOS distribution and the other does not. These
factors can all have a significant impact on census forecast accuracy
(see Littig and Isken 2007). Because different hospitals have different
methods for categorizing patients (different admitting services, DRGs
served, etc.), this requires a lengthy and ad-hoc procedure to be
performed at each new hospital, significantly impacting scalability.
For example, our industry co-author has indicated that this process
of clustering under current methods is unique to each hospital and
can take months to adequately determine patient types in large hospitals.

A second problem is that once the patient types have been identified,
trajectories are assigned based solely on what patient type the patient
is identified as. For example, if the patient is a bladder cancer
surgery patient their cluster will be bladder cancer surgery. However,
other factors that may impact the patient's trajectory and LOS, such
as age and gender, cannot be considered after the patient types are
defined. This approach is only as good as the granularity of each
cluster. However, the clusters are not defined based on the shape
of patients' location functions, but rather on other factors available
in the data that are believed to be associated with the shape of the
location function, but have not been statistically validated. Finally,
clusters cannot be too granular or data will be insufficient. This
phenomenon impacts both the ability to capture patient heterogeneity
and to accurately estimate patient paths because patients are forced
into predefined groups rather than assigned a type that most closely
matches their projected trajectory.

In contrast, we develop a new clustering approach that
clusters patients directly according to similarity of their trajectories
(which is what we want to estimate) in a statistically rigorous manner,
rather than using these ad-hoc proxies (e.g. DRG, age, gender). Specifically,
we seek to close the gap in the literature by developing new methods
for the CM step that provide more effective and scalable clustering
of patient types, and a better estimation of the patient trajectories
for each patient type. The proposed model, which we call \emph{clustering
and scheduling integration} (CSI) is scalable, captures the interactions
between hospital wards, and is capable of handling patient heterogeneity.
CSI begins with the CM module in which heterogeneous patients are
clustered based on the similarity of their trajectories. This provides
patient types for accurate estimation of patient trajectories and
patient census distributions at the ward level. Finally, these estimates
serve as inputs to the RS module to find an optimal hospital resource
schedule, which is then shown to outperform the same optimization
model using traditional CM methods.

For CM, we propose a novel semi-Markov mixture model (SMM) that integrates
the mixture clustering method and semi-Markov models accurately describing
stochastic location processes of patient trajectory. To the best of
our knowledge, this SMM clustering technique has not been proposed
before in the literature, either for the HASC problem or any other
problem. The SMM not only clusters patients based on their trajectory,
but also provides accurate estimates for the trajectory distribution
of each group of patients. In the RS module, the output of the CM
is fed into an MIP model similar to the model proposed by Helm and
Van Oyen (2015) to find the optimal resource schedule for hospitals.

We further show through a case study using real data from a partner
hospital that system performance is significantly impacted by the
quality of the input from the CM step. In fact, using CSI to parametrize
the optimization can enable up to a 50\% increase in elective admissions
while maintaining the same level of blocking and internal congestion
when compared with the same optimization using the traditional estimation
approach. Similarly, it is possible to have higher ward utilization
compared with traditional CM approaches holding all other metrics
constant.

The remainder of the paper is organized as follows. We first review
the literature and position the paper in Sec.~\ref{sec:lit}. Next,
we develop the new CSI methodology in Sec.~\ref{sec:CSI}, in which
the SMM clustering method for CM is discussed in detail, followed
by a brief description of the MIP model used for RS. Then in Sec.~\ref{sec:Validation-using-Simulation}
we use simulation to validate the proposed CSI model in terms of the
accuracy of estimates and the optimality of solutions. In Sec.~\ref{sec:Case-study-on},
we apply our CSI methodology in a case study based on historical data
from a partner hospital. Finally, in Sec.~\ref{sec:Conclusion-and-future}
we conclude the paper and discuss future opportunities.

\vspace{-0.2in}

\section{Literature}

\label{sec:lit}

\vspace{-0.2in}

Most existing research in the HASC area has focused on either CM or
RS separately. Little work can be found on integrating CM and
RS in a cohesive framework. Additionally, existing HASC approaches
lack at least one of the aforementioned characteristics of an effective
HASC model. The aim of this paper is to develop an HASC framework
that is scalable, accounts for patient heterogeneity, and considers
ward interactions through effective integration of CM and RS.

In the HASC literature, various stochastic and deterministic models
have been developed for RS. Green (2006) and Armony et al. (2011)
used queuing models to optimize resource scheduling.
Ward interactions were not taken into account in either of these papers.
Unlike the queuing models, simulation models developed for RS are
more flexible and consider the interaction between wards, mostly by
using patient pathways between wards in a hospital. Examples of simulation-based
models include Hancock and Walter (Hancock and Walter (1979, 1983)),
Griffith et al. (1976), Jacobson et al. (2006), Harper and Shahani
(2002), Zeltyn et al. (2011), and Konrad et al. (2013). However, simulation
models are case-specific, cannot be easily generalized or scaled,
and rely on the same, less effective PType and path estimation techniques
mentioned earlier. Adan et al. (2009), Bekker and Koeleman (2011),
and Zhang et al. (2009) used Mixed Integer Programming (MIP) models
for optimal RS. These works, however, only focus on either one ward
or an isolated feed-forward subset of the hospital, ignoring ward
interactions. To address this issue, Helm and Van Oyen (2015) proposed
a non-heuristic MIP scheduling model that also used patient pathways
to model ward interactions of an entire hospital. Although the RS
portion of the model is scalable and considers ward interactions,
it does not properly handle patient heterogeneity. Moreover, an empirical
method (similar to the traditional method described above) was used
to estimate the patient census at the ward level, which we show can
degrade the value of the optimal solution.

For RS optimization to be maximally effective, an accurate CM is required
to estimate patient arrival rates, their trajectory through the hospital,
and, by combining arrival and trajectory, the patient census at both
the ward and hospital levels. Regression analysis and time-series
modeling have been widely used for forecasting inpatient admissions
and hospital occupancy (Earnest et al. (2005) and Jones et al. (2002)).
Abraham et al. (2009) reviewed and compared several models for forecasting
daily emergency inpatient admissions and occupancy. They found that
the admissions are largely random and hence non-predictable, whereas
occupancy can be forecasted using a model combining regression and
ARIMA, or a seasonal ARIMA, for up to a week ahead. Their model is
capable of forecasting the overall hospital occupancy, but not the
occupancy at the ward level. Consequently, it does not account for ward
interactions. These approaches are also incapable of capturing
what-if scenarios or optimization with respect to inpatient admission
decisions. Littig and Isken (2007)
used occupancy flow equations to estimate occupancy at different units
or wards of a hospital. They predicted patient in- and out-flow using
time series and multinomial logistic regression models. They combined
these predictions and fed them into a set of flow equations to find
the net estimate of the number of patients in a given ward. However,
implementing this model in real time presents a major challenge, as
even a simple model requires coordination between a variety of real
time data sources and the computational burden of the method is high,
so scaling this model to large hospital would be difficult.

\begin{figure}
\begin{centering}
\begin{minipage}[t]{0.6\columnwidth}%
\subfloat[Phase-type model: Patients can transition in a sequential order or
leave the system\label{fig:Phase-type-model}]{\begin{centering}
\includegraphics[scale=0.31]{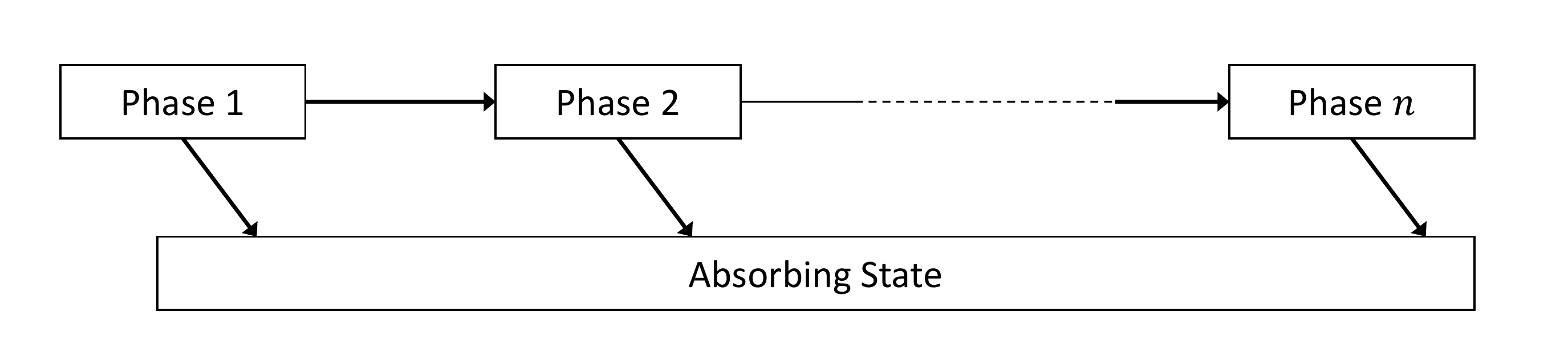} 
\par\end{centering}
}%
\end{minipage}
\par\end{centering}
\medskip{}

\begin{centering}
\begin{minipage}[t]{0.6\columnwidth}%
\begin{center}
\subfloat[Semi-Markov model: Any back and forth transition from any ward to
any ward is possible\label{fig:Semi-Markov-model}]{\begin{centering}
\includegraphics[scale=0.31]{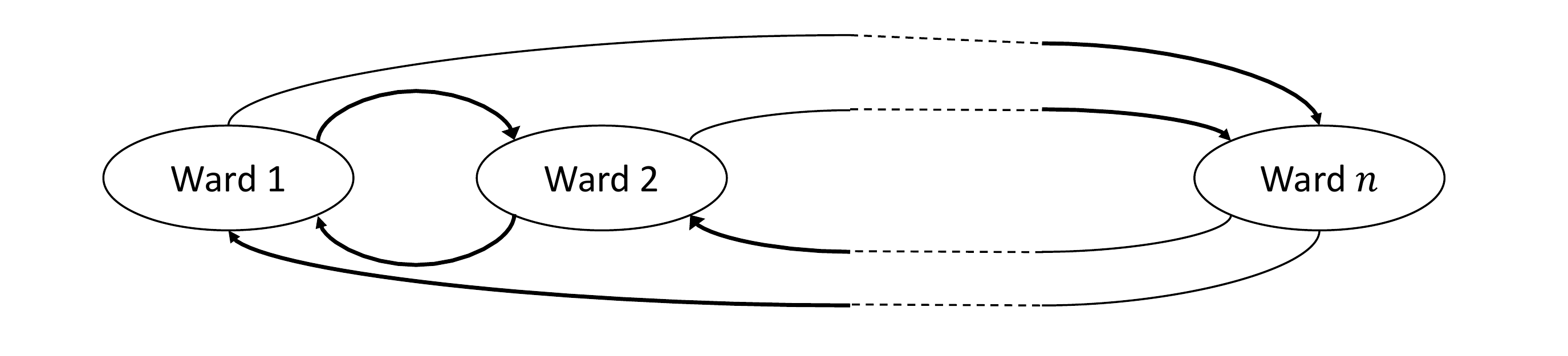} 
\par\end{centering}
}
\par\end{center}%
\end{minipage}
\par\end{centering}
\caption{Illustration of patient trajectory models for a hospital system}
\end{figure}

To model patient trajectory and LOS, Irvine et al. (1994) and Taylor
et al. (2000) proposed a continuous time Markov model for geriatric
patients. This model, however, was developed for few wards and lacks
scalability. Moreover, the assumption that the LOS at each ward follows
the same exponential distribution is not often a good model of reality.
Faddy and McClean (2000) used Phase-type distributions for patient
flow modeling. They interpreted phase-type distributions as a mixture
of components (phases) characterized by the severity of patient's
illness. Marshall and McClean (2003) extended this idea and developed
a model based on Conditional Phase-type distributions combined with
a Bayesian Network to be able to include a network of inter-related
variables representing causality. In phase-type methods, it is assumed
that the process begins in the first phase and may either progress
through the phases sequentially or enter an absorbing state (see Fig.~\ref{fig:Phase-type-model}).
Consequently, these methods cannot be extended to capture patient
trajectories, where patients revisit a ward several times or transition
from any ward to any other ward, which is a significant feature according
to our data. Thomas (1968) and Kao (1972, 1974) proposed a semi-Markov
model to predict recovery progress of coronary patients. This can
model any hospital system with complicated ward interactions in any
direction (See Fig.~\ref{fig:Semi-Markov-model}). Thus, this model
has \emph{scalability} and can fully model \emph{ward interactions}
but is built only for a ``homogenous'' mix of patients, i.e. coronary.

Patient heterogeneity is another challenge in CM. % and, consequently,
%patient trajectory estimation. 
To address this challenge, Helm and
Van Oyen (2015) partitioned patients into homogeneous clusters with
respect to their diagnosis using diagnosis related groups (DRGs).
DRGs have been also used by Fetter et al. (1980) for regional planning.
Harper (2005) provided a comprehensive review on clustering techniques,
including CART, k-means, neural network, etc. that use more patient
attributes (e.g., age, sex, diagnosis) to find more homogeneous clusters.
The main assumption of the DRG and attribute-based methods is that
patients who belong to a cluster, follow a similar trajectory. %and thus have similar expected services. 
However, this is not necessarily true.  Littig and Isken (2007) shows  
that, patients with similar attributes (e.g.,
age, sex, diagnosis, etc.) can often have very different trajectories. 
As an example from our own data, 
Fig.~\ref{fig:Case-Study-naive-trajectories} in Sec.~\ref{sec:Case-study-on} 
shows that although two patients shared the same age, sex, 
and diagnosis, their trajectories were very different.

In conclusion, the problem of trajectory estimation from a heterogeneous
cohort of patients is important. To our knowledge, existing
literature fails to address at least one or more challenges among:
scalability, ward interaction, and heterogeneity. In the next section,
we develop a methodology to address all three challenges and close this gap.

\vspace{-0.2in}

\section{Clustering and Scheduling Integrated (CSI) Model for HASC\label{sec:CSI}}

\vspace{-0.2in}

\begin{figure}
\begin{centering}
\includegraphics[scale=0.35]{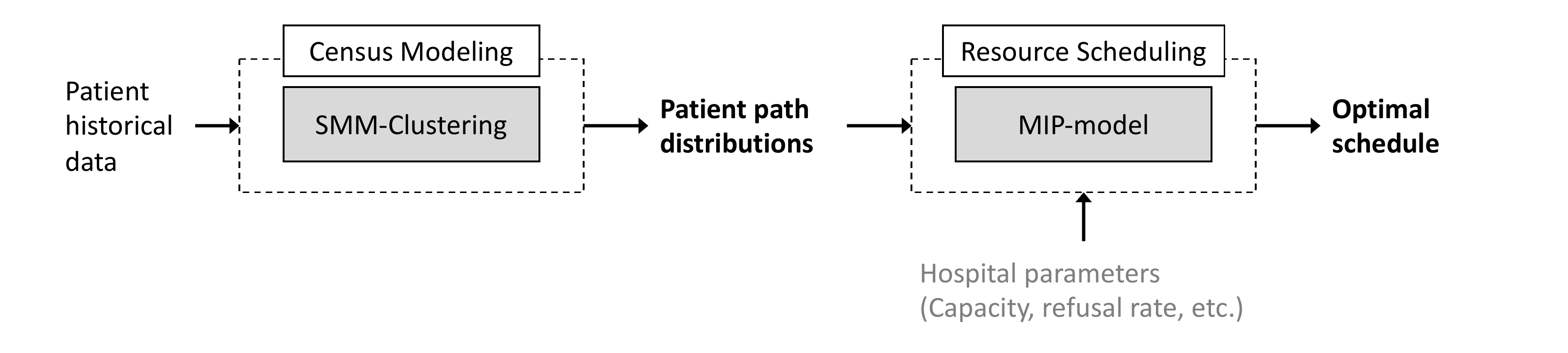} 
\par\end{centering}
\caption{Clustering and Scheduling Integrated (CSI) Model overview \label{fig:Methodology-overview}}
\end{figure}

Fig.~\ref{fig:Methodology-overview} provides a high level overview
of our methodology. First, historical patient flow data, taken from
admit-discharge-transfer (ADT) records, is used to group the patients
based on their trajectory using a semi-Markov Mixture (SMM) model-based 
clustering approach. The parameters for the semi-Markov processes
of patient trajectory for each cluster are estimated as a part of
the clustering process. These stochastic location processes are then
combined with a model of the non-stationary patient arrival process
to form a stochastic process (a Poisson arrival-location model or
PALM, see Massey and Whitt (1993)) that captures the ward-network
census. Estimation of this stochastic network census process
enables the derivation of three important products for hospital managers:
(1) Descriptive: accurate census forecasting, (2) What-if scenarios: 
impact of potential modifications to admission schedules, and (3)
Prescriptive: MIP-based admission scheduling optimization.

\subsection{Semi-Markov Mixture (SMM) Clustering for Modeling Patient Trajectories\label{subsec:Semi-Markov-Mixture-(SMM)}}

\vspace{-0.2in}

When a new patient arrives to the hospital, they are initially assigned
a bed in a hospital ward. The patient stays at that ward for a
stochastic duration and then transfers to another ward or
is discharged from the hospital. This process repeats if the patient
is transferred to another ward of the hospital.

A general hospital serves a cohort of many different types of patients.
Each type of patient requires different services during their hospital
stay. The first task is to identify patient types through clustering.
As mentioned previously, conventional clustering methods are not applicable
to this problem due to the fact two patients with the same observed
attributes often have different trajectories.

To manage the heterogeneous mix of patients in a hospital, we develop
a semi-Markov mixture model for clustering based on patient trajectory
rather than predefined groupings based on patient attributes. Patients
in each cluster are assumed to follow a semi-Markovian trajectory
through the hospital, which has been validated in the literature (e.g.
Hancock et. al. 1983). The SMM produces three important products that
significantly improve the generality and scalability of our method:
(1) appropriate patient groupings based on trajectory, (2) the optimal
number of patient types, and (3) accurate trajectories for each patient
type. In Sec.~\ref{sec:Case-study-on}, we show that this approach
yields more efficient patient clusters and more accurate trajectory
models than traditional approaches. Moreover, to the best of our knowledge, there is 
no existing approach for developing a semi-Markov mixture model and using it for clustering
spatial-temporal data. %In this section, we formulate a semi-Markov
%mixture model, and develop new EM-algorithm steps for clustering and
%estimating the semi-Markov process parameters.

\vspace{-0.2in}
\subsubsection{SMM Model Structure \label{subsec:Model}}

\vspace{-0.2in}

Let $\mathcal{K}$ be the set of unknown patient types, where each
patient type's trajectory follows a unique semi-Markov process. The
population of patient trajectory data, thus, follows a mixture of
an unknown number, $|\mathcal{K}|$, of semi-Markov processes.
Each mixture component, which we call a \emph{cluster} henceforth,
has a different semi-Markov process distribution. The first step is
to determine the set of clusters ( $\simeq\mathcal{K}$) and estimate
their corresponding trajectory distributions.

Consider a sample of trajectory data for $N$ patients observed over
a maximum time period of length $T$. Time is measured by discrete
units, for example, a day, quarter of day, hour, to be chosen depending
on the desired granularity. The set of possible lengths of stay
is denoted by $\mathcal{T} = \{1,2,\ldots,T\}$. 
Let $\mathcal{U}=\{\underline{\mathcal{U}},\bar{\mathcal{U}}\}$
denote the set of all states (wards) where $\underline{\mathcal{U}}$
is the set of all transient states and $\bar{\mathcal{U}}$ is the
set of all absorbing states. The first state when the patient enters
the system (hospital) is called the initial state and the last state,
which is an absorbing state, indicates a patient's end of stay in
the form of discharge or death. All the states during the patient's
hospital stay are transient states. The set of initial and transient
states are the same, as a patient may enter the hospital at any arbitrary
location.

A patient $n$'s ($n\in N$) trajectory is represented as ${\mathbf{y}^{(n)}}=(\{u_{1},\nu_{1}\},\ldots,\{u_{L^{(n)}},\nu_{L^{(n)}}\},\{\bar{u}\})$,
where $u_{l}\in\underbar{\ensuremath{\mathcal{U}}}$ indicates the
visited ward, $\nu_{l}\in\mathcal{T}$ is the length of stay at the
corresponding ward, $\bar{u}\in\bar{\mathcal{U}}$ is the absorbing
state from where the patient leaves the hospital, and subscript $l,\enspace l=1,2,..,L^{(n)},$
indicates the sequence of ward visits (\emph{state} and \emph{ward}
are used synonymously in this paper). $L^{(n)}$ is the patient $n$'s path 
sequence length. This model can capture general
network behavior, as there is no restriction on the number of times
a patient can visit any particular ward.

We formulate the problem by defining a set of parameters, $\mathbf{\Theta}=\{\Theta^{(k)}\},\:k\in\mathcal{K}$.
Each $\Theta^{(k)}$ is comprised of the mixture weight, $\pi^{(k)}$,
and semi-Markov process parameters, $\{\rho^{(k)},P^{(k)},H^{(k)}\}$,
for the $k$-th mixture. The mixture weight, $\pi^{(k)}$, denotes
the probability of a randomly chosen patient belonging to cluster $k$. Letting $z^{(n)}$
be a hidden variable representing the cluster index for patient $n$,
then the mixture weight can be expressed as, $\pi^{(k)}=p_{\mathbf{\Theta}}(z=k)$.
Also, $\sum_{k\in\mathcal{K}}\pi^{(k)}=1$.

Of the remaining mixture parameters, $\pmb{\rho}^{(k)}=\{\rho_{u}^{(k)}\},\,u\in\mathcal{U}$,
denotes the initial state probability. It can be expressed as $\rho_{u}^{(k)}=p_{\mathbf{\Theta}}(u_{1}=u|z=k)$,
the probability of the first state of a patient trajectory being ward $u$ given the patient belongs to cluster $k$. The matrix $\mathbf{P}^{(k)}=[P_{uj}],\,u,j\in\mathcal{U}$,
is the transition probability matrix, where $P_{uj}^{(k)}=p_{\mathbf{\Theta}}(u_{l}=j|u_{l-1}=u,z=k)$,
the probability of transitioning from ward $u$ to $j$ for a patient
in cluster $k$. Finally, $\mathbf{H}^{(k)}=[H_{uj}^{(k)}(\nu)],\,u,j\in\mathcal{U},\,\nu\in\mathcal{T}$,
is a three-dimensional tensor representing the holding mass distribution,
where $H_{uj}^{(k)}(\nu)=p_{\mathbf{\Theta}}(\nu_{l}=\nu|u_{l}=j,u_{l-1}=u,z=k)$
gives the probability of a patient in cluster $k$ spending $\nu$
time units in ward $u$ before transitioning to ward $j$ (from $u$).
As $\{\rho^{(k)},P^{(k)},H^{(k)}\}$ are probability distributions,
the following hold:

\vspace{-0.2in}

\begin{equation}
\sum_{u\in\mathcal{U}}\rho_{u}^{(k)}=1,\;\sum_{j\in\mathcal{U}}P_{uj}^{(k)}=1,\,and\,\sum_{\nu\in\mathcal{T}}H_{uj}^{(k)}(\nu)=1\label{eq:Prob-sum-1}
\end{equation}

\vspace{-0.2in}

\medskip{}

Using this parameterization, we represent the conditional probability
of any patient $n$'s trajectory, $\mathbf{y}^{(n)}$, given it is
generated by cluster $k$, in Eq.~\ref{eq:prob y given ck theta}.
The first part of the equation is the initial state probability. The
terms inside the product is the transition probability times the holding
time probability corresponding to the transition the patient made,
and the amount of time the patient spent at the ward before transitioning.
{\small{}{}{}{}{}{}{}}{\small \par}

{\small{}\vspace{-0.3in}
 {} 
\begin{eqnarray}
p_{\mathbf{\Theta}}(\mathbf{y}^{(n)}|z^{(n)}=k) & = & p(u_{1}|\rho^{(k)})\prod_{l=1}^{L^{(n)}}p(u_{l+1}|u_{l};\mathbf{P}^{(k)})p(\nu_{l}|u_{l+1},u_{l};\mathbf{H}^{(k)})\nonumber \\
 & = & \rho_{u_{1}}^{(k)}\prod_{l=1}^{L^{(n)}}\left\{ P_{u_{l},u_{l+1}}^{(k)}\cdot H_{u_{l},u_{l+1}}^{(k)}(\nu_{l}^{(i)})\right\} .\label{eq:prob y given ck theta}
\end{eqnarray}
}{\small \par}

{\small{}\vspace{-0.2in}
}{\small \par}

{\small{}\medskip{}
}{\small \par}

{\small{}{}}{\small \par}

\noindent Consequently, by considering the probability of belonging
to each cluster, $k$, the probability distribution function (pdf)
of the SMM model with $\mathcal{K}$ components is written as

\vspace{-0.3in}

{\small{}{}{}{}{}{}{ 
\begin{eqnarray}
p(\mathbf{y}^{(n)}|\mathbf{\Theta}) & = & \sum_{k\in\mathcal{K}}p_{\mathbf{\Theta}}(z^{(n)}=k)p_{\mathbf{\Theta}}(\mathbf{y}^{(n)}|z^{(n)}=k)\nonumber \\
 & = & \sum_{k\in\mathcal{K}}\pi^{(k)}\left[\rho_{u_{1}}^{(k)}\prod_{l=1}^{L^{(n)}}\left\{ P_{u_{l},u_{l+1}}^{(k)}\cdot H_{u_{l},u_{l+1}}^{(k)}(\nu_{l})\right\} \right].\label{eq:2-prob-y}
\end{eqnarray}
}}{\small \par}

{\small{}\vspace{-0.2in}
}{\small \par}

{\small{}\medskip{}
 {}}{\small \par}

\noindent Given an i.i.d. sample of $N$ patient trajectories, $\mathbf{Y}=\{\mathbf{y}^{(n)};n=1,\ldots,N\}$,
the likelihood function is, thus, given by

\vspace{-0.3in}

{\small{}{}{}{}{}{}{ 
\begin{eqnarray}
p_{\mathbf{\Theta}}(\mathbf{Y})=\prod_{n=1}^{N}p(\mathbf{y}^{(n)}\mathbf{|\mathbf{\Theta})=}\prod_{n=1}^{N}\sum_{k\in\mathcal{K}}\pi^{(k)}\left[\rho_{u_{1}}^{(k)}\prod_{l=1}^{L^{(n)}}\left\{ P_{u_{l},u_{l+1}}^{(k)}\cdot H_{u_{l},u_{l+1}}^{(k)}(\nu_{l})\right\} \right].\label{eq:3-prob-X-train}
\end{eqnarray}
}}{\small \par}

{\small{}\vspace{-0.2in}
}{\small \par}

{\small{}\medskip{}
 {}}{\small \par}

The parameters of the SMM mixture model, $\mathbf{\Theta}$, can be
estimated by maximizing the ($\log$)likelihood function in Eq.~\ref{eq:3-prob-X-train}.
However, if there is no observed transition between any two states
or no instance of any particular length of stay, the likelihood function
becomes zero. To avoid this issue, we use a Bayesian approach that
assigns very small prior probabilities to all model parameters, denoted
by $p(\mathbf{\Theta})$. Thus, according to Bayes rule, the posterior
probability for $\mathbf{\Theta}$ can be expressed as $p(\mathbf{\Theta}|{\bf Y})=\frac{p(\mathbf{Y}|\mathbf{\Theta})p(\mathbf{\Theta})}{p({\bf Y})}$.
Since $p({\bf {Y})}$ is independent of $\mathbf{\Theta}$, it suffices
to maximize the non-normalized posterior log-likelihood in Eq.~\ref{eq:MAP est}
to obtain the optimal $\mathbf{\Theta}^{*}$, also known as the $\emph{maximum a posteriori}$
(MAP) estimates of $\mathbf{\Theta}$.

\vspace{-0.3in}

{\small{}{}{}{}{}{}{ 
\begin{equation}
\mathbf{\Theta}^{*}=\arg\max_{\mathbf{\Theta}}\log\left\{ p({\bf Y}|\mathbf{\Theta})p(\mathbf{\Theta})\right\} \label{eq:MAP est}
\end{equation}
}}{\small \par}

{\small{}\vspace{-0.2in}
}{\small \par}

{\small{}\medskip{}
 {}}{\small \par}

The optimization problem in Eq.~\ref{eq:MAP est} does not have a
closed-form solution. %Hence, we develop an iterative \emph{expectation}-\emph{maximization} (EM) procedure in the following section to obtain the parameter estimates.Further,
Further, the non-normalized posterior log-likelihood function is non-convex 
 so Eq.~\ref{eq:MAP est} cannot
be solved using standard convex optimization methods. As a result,
we develop an iterative \emph{expectation}-\emph{maximization} (EM)
procedure in the following section to obtain the parameter estimates.

\vspace{-0.2in}
\subsubsection{Parameter Estimation via Expectation-Maximization (EM) \label{subsec:EM-Algorithm}}

\vspace{-0.2in}

An Expectation-Maximization (EM) algorithm is an effective approach
for learning maximum likelihood or maximum a posteriori (MAP) estimates,
where the likelihood is a function of unobserved latent variables
(in our case, $z$). It is an iterative approach comprising of an
Expectation (E-step) and Maximization (M-step) in each iteration.
In the E-step of any iteration $p$, we obtain a lower bound on the
objective function by taking its expectation at the current parameter
estimate, $\mathbf{\Theta}^{(p)}$. Then, in the M-step, we re-estimate
the parameters (update), to obtain $\mathbf{\Theta}^{(p+1)}$, that
maximizes the expectation from E-step. This procedure results in an
increase of the likelihood function with %and approaching a \textit{maxima }in each iteration. with
guaranteed convergence under some weak regularity conditions that
are satisfied in most practical situations (Wu, 1983). The specific
EM algorithm we develop for the SMM mixture model is as follows:

\subsubsection*{E-step}

\vspace{-0.2in}

We find the expected value of the maximum a posteriori function in
Eq.~\ref{eq:MAP est} with respect to the current parameter estimate,
$\mathbf{\Theta}^{(p)}$, denoted by $Q(\mathbf{\Theta}|\mathbf{\Theta}^{(p)})$
in Eq.~\ref{eq:Q-function-main}.

\vspace{-0.3in}

\begin{equation}
Q(\mathbf{\Theta}|\mathbf{\Theta}^{(p)})=\mathbb{E}_{\Theta^{(p)}}[\log(p({\bf Y}|\mathbf{\Theta})p(\mathbf{\Theta})]\label{eq:Q-function-main}
\end{equation}

\vspace{-0.2in}

\medskip{}

For a simpler expression of the $Q$ function in Eq.~\ref{eq:Q-function-main},
we define a \textit{membership }probability distribution. Membership
probability, denoted by $\Omega_{nk}$, is the probability of observing
any patient $n$'s trajectory, $\mathbf{y}^{(n)}$, generated by cluster
$k$, given parameters $\mathbf{\Theta}$ (see Eq.~\ref{eq:prik omega}).

\vspace{-0.3in}

\begin{eqnarray}
\Omega_{nk}(\mathbf{\Theta}) & = & \cfrac{\pi^{(k)}p_{\mathbf{\Theta}}(\mathbf{y}^{(n)}|z^{(n)}=k)}{\sum_{k'\in\mathcal{K}}\pi^{(k')}p_{\mathbf{\Theta}}(\mathbf{y}^{(n)}|z^{(n)}=k')}\label{eq:main-omega}\\
\Omega(\mathbf{\Theta}) & = & [\Omega_{nk}(\mathbf{\Theta})];\;n=1,\ldots,N,\,k\in\mathcal{K}\label{eq:prik omega}
\end{eqnarray}

\vspace{-0.2in}

\medskip{}

The $Q$ function, can thus be expressed as,

\vspace{-0.3in}

\begin{eqnarray}
Q(\mathbf{\Theta}|\mathbf{\Theta}^{(p)}) & = & \mathbb{E}_{\Theta^{(p)}}\left[\log(p({\bf Y}|\mathbf{\Theta})p(\mathbf{\Theta})\right]\nonumber \\
 & = & \sum_{n=1}^{N}\sum_{k\in\mathcal{K}}\Omega_{nk}(\mathbf{\Theta}^{(p)})\log\left[\pi^{(k)}p_{\mathbf{\Theta}}(\mathbf{y}^{(n)}|z^{(n)}=k)\right]+\log p(\mathbf{\Theta})\label{eq:Q function def for semi markov}
\end{eqnarray}

\vspace{-0.2in}

\medskip{}

\subsubsection*{M-step}

\vspace{-0.2in}

In the \emph{maximization} step, the parameters that maximize the
$Q$ function are estimated. The updated parameters are, thus,

\vspace{-0.3in}

\begin{equation}
\mathbf{\Theta}^{(p+1)}=\arg\max_{\Theta}\left\{ Q(\mathbf{\Theta}|\mathbf{\Theta}^{(p)})\right\} \label{eq:theta-p+1-est}
\end{equation}

\vspace{-0.2in}

\medskip{}

To solve Eq.~\ref{eq:theta-p+1-est}, we will estimate the posterior
of the parameters using a Dirichlet prior probability distribution
for $\mathbf{\Theta}$, $p(\mathbf{\Theta})$. The Dirichlet distribution
is chosen because 1) the parameters of a first-order semi-Markov mixture
are in the form of multinomial probabilities, which are suitably represented
by Dirichlet distribution, and 2) the conjugate of Dirichlet is also
a Dirichlet distribution, thus posterior computation is straightforward.

For any set of multinomial parameters, $x=(x_{1,}\ldots,x_{m})$,
such that $\sum_{i=1}^{m}x_{i}=1,\,0\leq x_{i}\leq1$, a Dirichlet
distribution is given by,

\vspace{-0.2in}

\begin{equation}
p(x_{1},\ldots,x_{m}|a_{1},\ldots,a_{m})=\cfrac{1}{B(a)}\prod_{i=1}^{m}x_{i}^{a_{i}-1}\label{eq:dirichlet}
\end{equation}

\vspace{-0.2in}

\medskip{}

\noindent where $a_{i}$'s are hyperparameters for $x$, and $B(a)=\cfrac{\prod_{i=1}^{m}\Gamma(a_{i})}{\Gamma\left(\sum_{i=1}^{m}a_{i}\right)}$,
a constant factor for the Dirichlet probability distribution function. 
Using the prior probability distributions, assumption of independence
of parameters, and plugging Eq.~\ref{eq:prob y given ck theta} into
Eq.~\ref{eq:Q function def for semi markov}, we obtain the posterior
distributions. We show in Online Appendix A, the posterior distributions
are Dirichlet, and how to update parameters to maximize Eq.~\ref{eq:Q function def for semi markov}.
The derived expressions are shown below,

\begin{eqnarray*}
\pi^{(k)} & = & \frac{\sum_{n=1}^{N}\Omega_{nk}(\mathbf{\Theta}^{(p)})+a_{\pi}^{(k)}}{\sum_{k'\in\mathcal{K}}\left[\sum_{n=1}^{N}\Omega_{nk'}(\mathbf{\Theta}^{(p)})+a_{\pi}^{(k')}\right]},\forall k\in\mathcal{K}.\\
\rho_{u}^{(k)} & = & \frac{\sum_{n=1}^{N}\Omega_{nk}(\mathbf{\Theta}^{(p)})\kappa(u_{1},u)+a_{\rho,u}^{(k)}}{\sum_{u'\in\mathcal{U}}\left[\sum_{n=1}^{N}\Omega_{nk}(\mathbf{\Theta}^{(p)})\kappa(u_{1},u')+a_{\rho,u'}^{(k)}\right]},\forall u\in\mathcal{U},k\in\mathcal{K}
\end{eqnarray*}
\begin{eqnarray*}
P_{uj}^{(k)} & = & \frac{\sum_{n=1}^{N}\Omega_{nk}(\mathbf{\Theta}^{(p)})\bar{\kappa}_{uj}(\mathbf{y}^{(n)})+a_{P,uj}^{(k)}}{\sum_{j'\in\mathcal{U}}\left[\sum_{n=1}^{N}\Omega_{nk}(\mathbf{\Theta}^{(p)})\bar{\kappa}_{uj'}(\mathbf{y}^{(n)})+a_{P,uj'}^{(k)}\right]},\forall u,j\in\mathcal{U},k\in\mathcal{K}\\
H_{uj}^{(k)}(\nu) & = & \frac{\sum_{n=1}^{N}\Omega_{nk}(\mathbf{\Theta}^{(p)})\tilde{\kappa}_{uj,\nu}(\mathbf{y}^{(n)})+a_{H,uj}^{(k)}(\nu)}{\sum_{\nu'\in\mathcal{T}}\left[\sum_{n=1}^{N}\Omega_{nk}(\mathbf{\Theta}^{(p)})\tilde{\kappa}_{uj,\nu'}(\mathbf{y}^{(n)})+a_{H,uj}^{(k)}(\nu')\right]},\forall u,j\in\mathcal{U},\nu\in\mathcal{T},k\in\mathcal{K}
\end{eqnarray*}

\subsubsection{SMM-Clustering Algorithm\label{subsec:SMM-Clustering-Algorithm}}

In this section, we detail the specific algorithm for implementing SMM clustering 
(see Algorithm~\ref{alg:SMM-clustering-algo}) and discuss key features such as 
sensitivity to initialization, identifiability, computational complexity, 
and optimization acceleration techniques.

\begin{figure}
\noindent\begin{minipage}[t]{1\columnwidth}%
\begin{algorithm}[H]
\small
	\caption{SMM-Clustering Algorithm}
	\label{alg:SMM-clustering-algo}
	\begin{algorithmic}[0]
	\INPUT Trajectory data, $\mathbf{Y}=\{\mathbf{y}^{(n)};n=1,\ldots,N\}$, number of clusters, $K$.
	
		\Initialize{$z^{(n)}\gets \text{rand}(1,K),n=1,\ldots, N.$
		\Comment{s.t. at least one trajectory assigned to each $k$}\\
		$\Omega \gets \{1/K\}_{N \times K} $
		
		$a_\pi \gets \epsilon/K$; 
		$a_\rho \gets \frac{\epsilon}{|\mathcal{U}| \times K}$; \\
		$a_\mathbf{P} \gets \frac{\epsilon}{|\mathcal{U}| \times |\mathcal{U}| \times K}$;
		$a_\mathbf{H} \gets \frac{\epsilon}{|\mathcal{U}| \times |\mathcal{U}| \times |\mathcal{U}| \times K}$
		\Comment{Prior hyperparameters}}
			
			\For{$\text{iter} = 1,\ldots, \text{maxIter}$}
				\State {$\boldsymbol{\Theta} \gets$ \Call{SMMParameters}{$\Omega,\mathbf{z}$}}
				\State {$\Omega \gets $ \Call{MembershipProb}{$\boldsymbol{\Theta},\mathbf{z}$}}
				\State {$z^{(n)} \gets \arg \max_{k} \Omega_{n,k}, n=1,\ldots,N$}
			\EndFor

	\Function{MembershipProb}{$\boldsymbol{\Theta},\mathbf{z}$}
		\State $\Omega \gets \mathbf{0}_{N \times K}$
		\For{$k = 1,\ldots,K$}
			\For{$n = 1,\ldots,N$}
				\State {Fetch trajectory sequence, ${\mathbf{y}^{(n)}}=(\{u_{1},\nu_{1}\},\ldots,\{u_{L^{(n)}},\nu_{L^{(n)}}\},\{\bar{u}\})$}
				\State {$\Omega_{n,k} \gets \pi^{(k)}*\rho^{(k)}_{u_1}$}
				\For{$l = 1,\ldots,L^{(n)}$}
					\State {$\Omega_{n,k} \gets \Omega_{n,k}*\mathbf{P}^{(k)}_{u_l,u_{l+1}}*\mathbf{H}^{(k)}_{u_l,u_{l+1}}(\nu_l)$}
				\EndFor
			\EndFor
		\EndFor

		\State {$\Omega_{n,k} \gets \frac{\Omega_{n,k}}{\sum_{k'=1}^{K}\Omega_{n,k'}}; k=1,\ldots,K, n=1,\ldots,N $} \Comment{Normalizing for, $\sum_{k=1}^{K}\Omega_{n,k}=1,\forall n=1,\ldots,N$}

		\State \Return {$\Omega$}
	\EndFunction

	\Function{SMMParameters}{$\Omega,\mathbf{z}$}
		\State {$\pi^{(k)} \gets a_\pi + \sum_{n=1}^N \Omega_{n,k}; k=1,\ldots, K$}
		\State {$\pi^{(k)} \gets \pi^{(k)}/\sum_{k'=1}^K\pi_{k'}$} \Comment{Normalizing}

		\State {$\rho \gets \mathbf{0}_{\mathcal{U} \times K}; \mathbf{P} \gets \mathbf{0}_{\mathcal{U} \times \mathcal{U} \times K}; \mathbf{H} \gets \mathbf{0}_{\mathcal{U} \times \mathcal{U} \times \mathcal{T} \times K}$}
		\State {$\rho^{(k)}_u \gets a_\rho; u \in \underbar{$\mathcal{U}$}, k=1,\ldots,K$}
		\State {$\mathbf{P}^{(k)}_{u,u'} \gets a_\mathbf{P}; u,u'\in \underbar{$\mathcal{U}$}, u \neq u', k=1,\ldots,K$}
		\State {$\mathbf{H}^{(k)}_{u,u'}(\nu) \gets a_\mathbf{H}; u,u'\in \underbar{$\mathcal{U}$}, u \neq u', \nu \in \mathcal{T}, k=1,\ldots,K$}

		\For{$n = 1,\ldots,N$}
			\State {Fetch trajectory sequence, ${\mathbf{y}^{(n)}}=(\{u_{1},\nu_{1}\},\ldots,\{u_{L^{(n)}},\nu_{L^{(n)}}\},\{\bar{u}\})$}
			
			\State {$\rho_{u_1}^{z^{(n)}} \gets \rho_{u_1}^{z^{(n)}} + \Omega_{n,z^{(n)} }$}
			
			\For{$l = 1,\ldots,L^{(n)}$}
				\State {$\mathbf{P}_{u_l,u_{l+1}}^{z^{(n)}} \gets \mathbf{P}_{u_l,u_{l+1}}^{z^{(n)}} + \Omega_{n,z^{(n)}}$}

				\State {$\mathbf{H}_{u_l,u_{l+1}}^{z^{(n)}}(\nu_l) \gets \mathbf{H}_{u_l,u_{l+1}}^{z^{(n)}}(\nu_l) + \Omega_{n,z^{(n)}}$}
			\EndFor
		\EndFor
		\Comment {Normalizing as per Eq.~\ref{eq:Prob-sum-1}}
		\For{$k = 1,\ldots, K$}
			\State {$\rho^{(k)}_u \gets \rho^{(k)}_u/\sum_{u'\in \mathcal{U}}\rho^{(k)}_{u'}; \forall u\in \mathcal{U}, k=1,\ldots,K$}
			\State {$\mathbf{P}^{(k)}_{u,j} \gets \mathbf{P}^{(k)}_{u,j}/\sum_{j'\in \mathcal{U}}\mathbf{P}^{(k)}_{u,j'}; \forall u,j\in \mathcal{U}, k=1,\ldots,K$}
			\State {$\mathbf{H}^{(k)}_{u,j}(\nu) \gets \mathbf{H}^{(k)}_{u,j}(\nu)/\sum_{\nu'\in \mathcal{T}}\mathbf{H}^{(k)}_{u,j}(\nu'); \forall u,j\in \mathcal{U}, \nu \in \mathcal{T}, k=1,\ldots,K$}
		\EndFor

		\State \Return {$\boldsymbol{\Theta} = \{\pi^{(k)}, \rho^{(k)}, \mathbf{P}^{(k)}, \mathbf{H}^{(k)}; k=1,\ldots,K\}$}
	\EndFunction

\OUTPUT {$\hat{\mathbf{\Theta}}=\{\hat{\Theta}^{(k)}\},k=1,\ldots,K$; and cluster assignments $\mathbf{z}.$}

	\end{algorithmic}
\end{algorithm}%
\end{minipage}
\end{figure}

As shown in Algorithm~\ref{alg:SMM-clustering-algo}, we take the
trajectory data and the number of clusters as inputs. The estimation
procedure is initialized by randomly assigning a cluster to each patient,
such that each cluster in $1,\ldots,K$ has at least one patient.
The membership probabilities in $\Omega$ are initialized
as uniform (any other random assignment can also be done). The Dirichlet
prior hyperparameters, $a_{(\cdot)}$, are chosen as a small number
and uniform for all parameters ($\epsilon$ is a small positive number,
taken as $1e-5$ in our experiments). Thereafter, iterative estimation
is done, where the membership probabilities and the SMM parameters
are updated in each iteration.

In our implementation, we set a termination condition so that 
the algorithm terminates after $\text{maxIter}$ 
iterations. Iterations can also be performed
until a given measure of convergence. For example, convergence can
be measured in terms of either no change in the objective function
or hard cluster assignments of the trajectories ($\mathbf{z}$) -- i.e. 
the clusters do not change much between iterations. Tracking
of cluster reassignments (in each iteration) works better than tracking
the objective function, as the change in latter becomes extremely
small after few iterations. But, in practice, having an upper bound
for the number of iterations ($\text{maxIter}$) is more useful due
to potential identifiability issues. Especially when the data size
is large, it can take a very long time for cluster reassignments of 
all trajectories to stabilize between iterations. $\text{maxIter}$ serves
as a reasonable trade-off between computation time and accurate results,
and hence is commonly employed in many clustering implementations.

The runtime computational complexity of the algorithm is linear in 
the length of sequences, the sample size, the number of clusters,
and the number of iterations, i.e. $O(\text{maxIter}*KNL)$, where
$L$ is average sequence length. This linear complexity makes the implementation
fast. Additionally, several steps in the algorithm can be vectorized, e.g.
parameter normalization, for increased speed. Computation time 
can be further reduced by: (1) \emph{parallelization:} since the parameter update
equations are independent for each cluster, state, and length-of-stay,
we can split the computation across many computing nodes; and (2) stochastic
clustering: using a random subsample of data for parameter updating 
in each iteration. Parallelization requires multiple
computing nodes, while the stochastic clustering is suitable when the data sample
is very large. A higher order Markov clustering extension of our proposed 
model will increase the computational complexity, and therefore may 
require some or all of the above techniques for tractable solution times.

Similar to most other clustering methods, the SMM-clustering results
are sensitive to the initialization. While some experts suggest using 
 prior system knowledge for initialization, such as diagnosis-related-groups 
(DRG), we feel such an initialization may introduce bias into our results; particularly  because 
a main motivation for developing this approach was that DRG clustering 
was not sufficiently accurate in practice. 
We, therefore, recommend random cluster initialization. To avoid potentially  
poor solutions resulting from a particular initialization, we perform multiple 
runs with different random initializations and choose the solution 
with the highest final objective function.

\vspace{-0.2in}
\subsubsection{Determining the number of clusters \label{subsec:Choice-of-clusters}}

\vspace{-0.2in}

To determine the appropriate number of clusters, we estimate the SMM
model and compute the $Q$ function, which is analogous to the likelihood.
We then increase the number of clusters, $|\mathcal{K}|$, by one at each iteration.  
We stop when there is no significant change in the $Q$ function by adding 
an additional cluster (popularly known
as the \textit{elbow }method). To eliminate redundant clusters, 
we perform pairwise hypothesis tests with controlled type-I
error for the identified clusters. We use the Chi-square hypothesis
test developed by Billingsley (1961a, 1961b) for comparing transition
probabilities and Kolmogorov-Smirnov for comparing the distributions
on the initial state and the holding time. We merge any clusters that
are found similar by these tests and then perform the tests again
in iterative fashion until no redundant clusters are detected. A similar
approach for removing redundant clusters was used by Weiss et. al
(1982).
\vspace{-0.2in}
\subsubsection{Trajectory estimation for each cluster \label{subsec:Trajectory-estimation-for}}

\vspace{-0.2in}

After parameter estimation, the next step is to estimate the patient
trajectory distributions which are characterized by the visited wards
and length of stay at each ward. Using the selected number of clusters
and corresponding semi-Markov process estimates from our EM algorithm, 
we compute the probability distribution of patient trajectory,
denoted by $\mathbf{\Gamma}(d)=[\gamma_{j}^{(k)}(d)];\enspace j\in\mathcal{U},k\in\mathcal{K}$
and $d=1,2,\ldots$, where $\gamma_{j}^{(k)}(d)$ is the probability
that a patient of cluster $k$ is in ward $j$ after $d$ days (we
use a day as a time unit for $\nu$). This distribution is one of the
key inputs to the scheduling optimization.

To estimate $\mathbf{\Gamma}(d)$ we use \emph{interval transition probabilities},
$\Phi^{(k)}=[\phi_{uj}^{(k)}(d)];\enspace u,j\in\mathcal{U},k\in\mathcal{K}$
and $d=1,2,\ldots$, where $\phi_{uj}^{(k)}(d)$ is the probability
that a patient in cluster $k$ is in ward $j$ on day $d$, given
that the patient entered the hospital in ward $u$. Recalling that,
for a type $k$ patient, $H_{uj}^{(k)}(d)$ is the holding time probability
distribution in ward $u$ before transitioning to ward $j$ and $P_{uj}^{(k)}$
is the probability of transitioning from ward $u$ to $j$, then $\phi_{uj}^{(k)}(d)$
is computed as

\vspace{-0.2in}

{\small{}{}{}{}{}{}{ 
\begin{flalign}
\phi_{uj}^{(k)}(d) & =P_{uj}^{(k)}H_{uj}^{(k)}(d)+\delta_{uj}\sum_{l\in\mathcal{U}\backslash\{u\}}\sum_{d'=d+1}^{\infty}P_{ul}^{(k)}H_{ul}^{(k)}(d')+\sum_{l\in\underline{\mathcal{U}}\backslash\{j\}}\sum_{d'=1}^{d}P_{ul}^{(k)}H_{ul}^{(k)}(d')\phi_{lj}^{(k)}(d-d'),\label{eq:phi-i-j-new}
\end{flalign}
}}{\small \par}

\vspace{-0.2in}

\medskip{}

\noindent where $\delta_{uj}=\begin{cases}
1, & u=j\\
0, & u\neq j
\end{cases}$ and $\phi_{uj}^{(k)}(0)=\begin{cases}
1, & u=j\\
0, & u\neq j
\end{cases}$. A patient starting in state $u$ can be in state $j$ on day $d$
either if the patient stays in ward $u$ for $d$ days before transitioning
to ward $j$ (the first term of Eq.~\ref{eq:phi-i-j-new}), or $u=j$
and they never left $u$ during the period $[0,d]$ (the second term
of Eq. \ref{eq:phi-i-j-new}), or the patient left $u$ at least once
and finally reached $j$ by day $d$ (the third term of Eq.~\ref{eq:phi-i-j-new}).
Consequently, $\gamma_{j}^{(k)}(d)$ can be expressed as sum-product
of all possible initial states to ward $j$ (Eq.~\ref{eq:gamma-input to schedule}).

\vspace{-0.3in}

\begin{flalign}
\gamma_{j}^{(k)}(d) & =\sum_{u\in\mathcal{U}}\rho_{u}^{(k)}\phi_{uj}^{(k)}(d)\label{eq:gamma-input to schedule}
\end{flalign}

\vspace{-0.15in}

\noindent $\gamma$ from Eq. \ref{eq:gamma-input to schedule} becomes an input to the \emph{scheduling}
model explained in next section. The semi-Markov process estimates,
$\hat{\mathbf{\Theta}}$, can be used for finding the length-of-stay
distribution of each patient type as well as the expected mean length
of stay in each ward and its variance. Equations to compute these
are given in the following subsection as they may be useful for other
research objectives or purposes.
\vspace{-0.2in}
\subsubsection{Computing Patient length-of-stay distributions \label{subsec:Appendix-B:-Patient LOS}}
\vspace{-0.15in}
\noindent \textbf{Length-of-stay in a ward ($V$).} For a patient of type $k$, we estimate the expected days spent by
the patient in each ward using the indicator function on the interval transition probability $\Phi^{(k)}$. 
Let $\bar{V}^{(k)}=\left[\bar{v}_{uj}^{(k)}\right];\:u,j\in\mathcal{U};\:k\in\mathcal{K}$,
where $v_{uj}^{(k)}$ denotes the number of days the patient will
spend in $j$ given their initial state was in ward $u$. The mean
of $v_{uj}^{(k)}$ can be computed using Eq. \ref{eq:v-bar-i-j} given
below.

\vspace{-0.4in}

\begin{eqnarray}
\bar{v}_{uj}^{(k)} & = & \sum_{d=1}^{\infty}\phi_{uj}^{(k)}(d)\label{eq:v-bar-i-j}
\end{eqnarray}

\vspace{-0.2in}

The second moment of $v_{uj}^{(k)}$is given by

\vspace{-0.4in}

\begin{eqnarray}
\bar{v}_{uj}^{2(k)} & = & \bar{v}_{uj}^{(k)}(2\bar{v}_{uj}^{(k)}-1)\label{eq:v-i-j-2nd}
\end{eqnarray}

\vspace{-0.2in}

\noindent Thus, the variance of the days spent by a patient in a state can be
given by

\vspace{-0.4in}

\begin{eqnarray}
\check{v}_{uj}^{(k)} & = & \bar{v}_{uj}^{2(k)}-(\bar{v}_{uj}^{(k)})^{2}\quad\forall u,j\in\mathcal{U}.\label{eq:v-var-i-j}
\end{eqnarray}

\vspace{-0.2in}

\noindent \textbf{Total hospital length-of-stay (LOS).} To get the distribution 
on LOS for entire hospital stay, we 
calculate the first-passage-time probabilities, denoted by $F$. 
$F^{(k)}=\left[f_{uj}^{(k)}(d)\right],\:u,j\in\mathcal{U};\:\nu=1,2,\ldots;\:k\in\mathcal{K}$,
where $f_{uj}^{(k)}(d)$ is the probability that the first passage
from state $u$ to $j$ will take exactly $d$ days for patients of
type $k$. This event can occur if a patient makes a direct transition
from $u$ to $j$ on day $d$, or the patient transitions to any other
state $l$ on any day before $d$ and then takes first passage from
$l$ to $j$. The second component is recursive and thus takes into
account any number of transitions between any states (except the absorbing
state) to reach state $j$ from $u$ in $d$ days.

\vspace{-0.4in}

\begin{eqnarray}
f_{uj}^{(k)}(d) & =P_{uj}^{(k)}H_{uj}^{(k)}(d)+\sum_{l\in\underline{\mathcal{U}}\backslash\{j\}}\sum_{d'=1}^{d}P_{ul}^{(k)}H_{ul}^{(k)}(d')f_{lj}^{(k)}(d-d').\label{eq:f-ij}
\end{eqnarray}

\vspace{-0.2in}

\noindent Using $f_{u\bar{u}}^{(k)}$, where $\bar{u}\in\mathcal{\bar{U}}$, and the initial
state probability $\mathbf{\rho}$ we can get the distribution for
LOS. $f_{u\bar{u}}^{(k)}$ denotes the first-passage-probability for a patient's
flow from any initial state $u$ to a discharge state $\bar{u}$. If the
initial state is unknown then we use Eq.~\ref{eq:L-n-LOS}. Otherwise
if the initial state is known, say $u$, then the distribution is
given by $f_{u\bar{u}}^{(k)}$ itself.

\vspace{-0.4in}

\begin{eqnarray}
L^{(k)}(d) & = & \sum_{u\in\underbar{\ensuremath{\mathcal{U}}}}\rho_{u}^{(k)}f_{u\bar{u}}^{(k)}(d)\quad d=1,2,\ldots\label{eq:L-n-LOS}
\end{eqnarray}

\vspace{-0.2in}

\subsubsection{Elective and emergency inpatient census model \label{subsec:SMM-Elective-and-Emergency}}
\vspace{-0.2in}
In this section, we describe how we integrate the semi-Markov stochastic
location processes generated from our SMM method with different arrival
processes to create a stochastic ward census process. This section,
as well as Sec.~\ref{subsec:Mixed-Integer-Programming}, presents
an elective scheduling optimization approach focused on hospital patient
throughput (i.e. admission volume) and congestion (e.g. bed block,
off-ward placement of patients) that is based on the work by Helm
and Van Oyen (2015). The purpose of these sections is to provide relevant
background for possible applications of our CM method in the hospital
census forecasting industry as described by our industry co-author.
We use the aforementioned optimization approach as a proof of concept
to test the value of our improved CM method and demonstrate how our CM 
approach integrates seamlessly with existing patient flow optimizations. 
These sections are, therefore,
intentionally brief and not intended to present new research in the
area of resource optimization. %The focus of this paper is on the development
%and analysis of a CM method (patient type clustering and trajectory
%estimation) that is designed to integrate with existing optimization
%approaches such as the one presented herein.

There are two broad categories of patients that a hospital serves,
elective (EL) and emergency (EM). In developing our census model we
separate the two because in the optimization in Sec.~\ref{subsec:Mixed-Integer-Programming},
emergency arrivals are considered uncontrollable while the scheduled
elective arrivals become the primary decision variable. To integrate
our SMM clustering and trajectory estimates with the optimization
as well as the what-if scenarios of interest to the industry, we run
the clustering method on EL and EM patients separately. Hence each
stream, EL and EM, will have its own set of patient types, $\mathcal{K}$,
with their own trajectories determined by our SMM.

As explained in previous sections (\ref{subsec:Model}-\ref{subsec:Choice-of-clusters}),
we cluster the EL patients into homogeneous groups with similar trajectories.
Trajectory estimates, one for each patient \emph{type} (cluster),
are computed using Eqs.~\ref{eq:phi-i-j-new} and \ref{eq:gamma-input to schedule}.
Combining the EL arrival pattern with the semi-Markov trajectory distributions
for each patient type, discussed in Sec.~\ref{subsec:Trajectory-estimation-for},
creates a stochastic census process that can be used to calculate
the distribution on patient demand for beds at each ward at any time,
$t$. The exact distribution depends on the arrival process.

For EL admissions we consider a deterministic arrival process, which,
when combined with the semi-Markovian patient trajectories, yields
a Poisson-Binomial distribution on bed demand at fixed time point
$t$. The deterministic assumption is an approximation of reality,
but has been widely used in the literature due to the fact that elective
arrivals are controlled and scheduled in advance. Therefore it is
(1) theoretically possible to achieve close to a deterministic arrival
stream, (2) it is highly beneficial to patient flow for hospital managers
to work toward a deterministic elective arrival stream and should
be a management priority, (3) deviations from the deterministic arrivals
can be incorporated for certain distributions and approximated for
others \textemdash{} particularly if the variance of the arrival pattern
can be adequately approximated as a linear function of the mean.

We model the arrivals of emergency patients using a non-homogeneous Poisson
process that varies by day of week. Combining these Poisson arrivals
with the semi-Markov stochastic location processes yields a Poisson-arrival-location
model (PALM) of emergency census, (see Massey and Whitt (1993) for
more details). One feature of a PALM model is that the distribution
on demand for beds in any ward for fixed $t$ follows a Poisson distribution.

Having defined the distribution on demand for beds for emergency and
elective patients, we now briefly describe an optimization model from
the literature (Helm and Van Oyen (2015)) that is subsequently used
to demonstrate the importance of a rigorous patient trajectory estimation
procedure. We designed our estimation approach to integrate with optimization
and what-if scenarios, with this particular optimization being used
as a proof of concept that (1) our method integrates well with current
optimization approaches, and (2) our method significantly improves
the outcome of the optimization when compared with traditional approaches
proposed for use with these types of models.

\vspace{-0.2in}
\subsection{Resource Scheduling (RS) MIP model for Elective Admission Scheduling
\label{subsec:Mixed-Integer-Programming}}

\vspace{-0.2in}

%As mentioned above (Sec.~\ref{subsec:SMM-Elective-and-Emergency}),
%the schedule of EL admissions can be controlled while EM arrivals
%are not in hospital's control. 
The RS model we use as proof of concept
integrates both EM and EL census models to capture metrics such as
blocking and off-ward placement of patients. The two common objectives
from the literature that we focus on are: 1) maximizing the number
of elective admissions while constraining congestion metrics and 2)
minimizing the congestion (e.g. blocking) while maintaining patient
throughput. From a management perspective, the first objective allows
for increased revenue, while the second objective provides better
access and consequently better outcomes for patients. For ease of
reference, we present this optimization model in Online Appendix B.
%\ref{app:elecopt}.

%\bigskip{}

This concludes the presentation of our CSI approach. %that develops
%an optimal design of patient types and trajectory estimations for
%integration into an inpatient admission scheduling optimization model.
In the next section we develop a simulation to validate the accuracy
of the SMM approach for patient clustering and trajectory estimation
and to determine the impact of the SMM on optimal solutions to the
MIP model.

\vspace{-0.2in}

\section{SMM Validation and Impact on Optimal Scheduling Solutions\label{sec:Validation-using-Simulation}}

\vspace{-0.2in}

In this section, we perform simulation studies to validate the performance
of SMM method. 
\vspace{-0.2in}
\subsection{Evaluating the accuracy of the SMM method}

\vspace{-0.2in}

\begin{figure}
\begin{centering}
\begin{minipage}[c]{0.45\columnwidth}%
\begin{center}
\subfloat[Data generating transition probabilities\label{fig:Data-generating-transition}]{\begin{centering}
\includegraphics[scale=0.42]{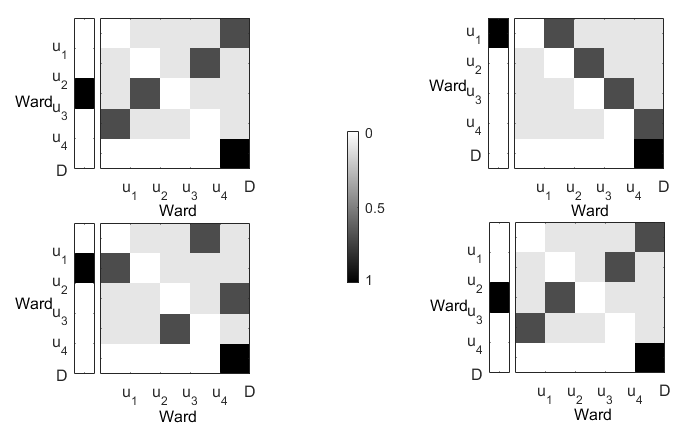} 
\par\end{centering}
}
\par\end{center}%
\end{minipage}\hfill{}%
\begin{minipage}[c]{0.45\columnwidth}%
\subfloat[Estimated cluster's transition probabilities\label{fig:Computed-clusters'-transition}]{\begin{centering}
\includegraphics[scale=0.42]{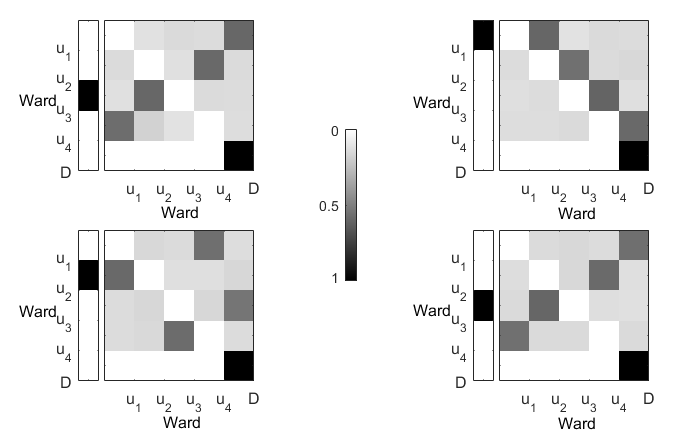} 
\par\end{centering}
}%
\end{minipage}
\par\end{centering}
\caption{Pictorial representation of transition probabilities as a gray scale
heat-map; with higher intensity of gray for higher probability. The
heat-map for generating and estimated cluster transition probabilities
are shown side-by-side for visual comparison.}
\end{figure}

We begin with a detailed analysis of the functionality and performance of 
our SMM method by performing a simulation study of a hospital system with four transient states (wards), 
$\underline{\mathcal{U}}=\{u_{1},\ldots,u_{4}\}$ and one
absorbing state (discharge/death) $\bar{\mathcal{U}}=\{D\}$. 
We later expand upon this deep-dive to consider a variety of other clustering 
systems. For the initial simulation, 
flow sequences for 1000 patients were generated from four different
semi-Markov models (corresponding to four different patient types),
denoted by $C_{s}^{(1)},\ldots,C_{s}^{(4)}$. As two clusters could
be different in $\mathbf{P}$, $\mathbf{H}$, and/or both, we used
the following setting that covers all possible scenarios. In the data
generating model, $C_{s}^{(1)}$ and $C_{s}^{(2)}$ have different
$\mathbf{P}$ but same $\mathbf{H}$, $C_{s}^{(3)}$ and $C_{s}^{(4)}$
have same $\mathbf{P}$ and different $\mathbf{H}$, while $C_{s}^{(2)}$
and $C_{s}^{(3)}$ have different $\mathbf{P}$ and $\mathbf{H}$.
A pictorial representation of the transition probability matrix combined
with the initial state probability is shown in Fig.~\ref{fig:Data-generating-transition}.
In these plots, the darker the color, the higher the probability.
The component mixture weights, $\mathbf{\pi}$, of the four clusters
are $\{0.17,0.33,0.25,0.25\}$ respectively. Additionally, the assignment
probabilities in the generating distributions were set less than 0.7
to ensure that the simulation output would be similar to that of a
general hospital scenario.

\begin{figure}[h]
\begin{centering}
\includegraphics[scale=0.7]{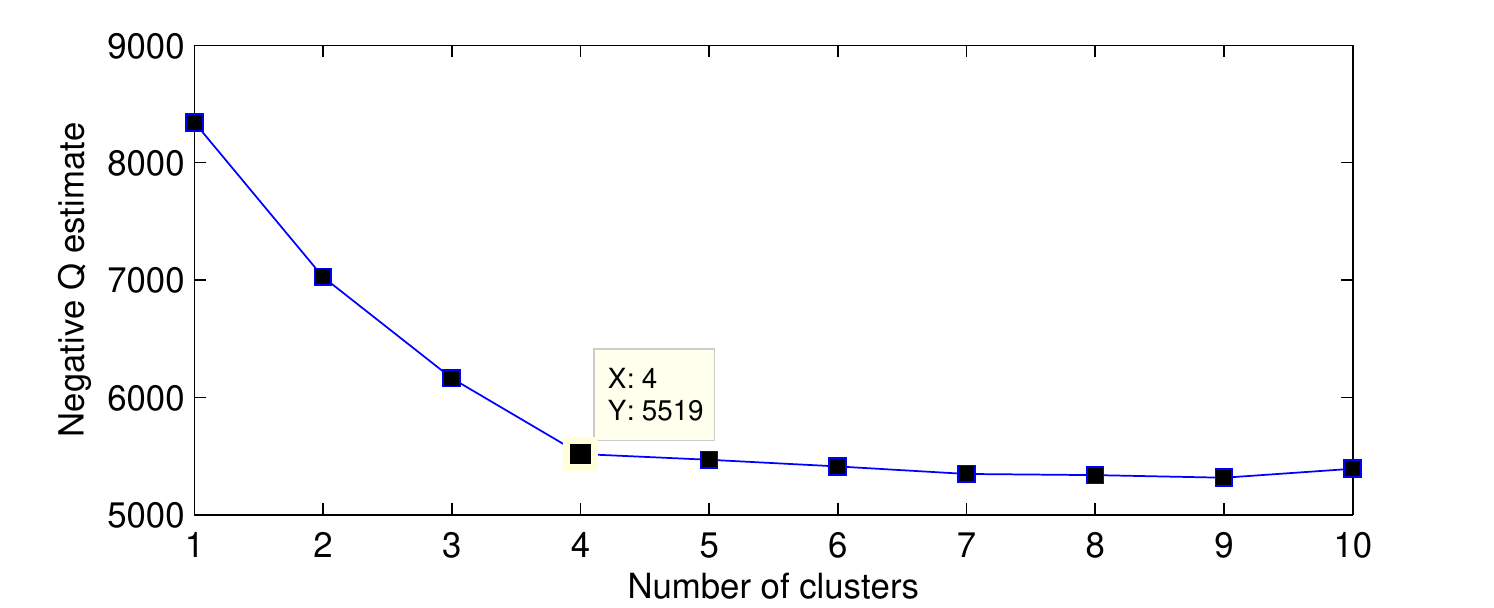} 
\par\end{centering}
\caption{Q function estimates against number of clusters for the simulated
data. The improvement in Q estimate becomes insignificant after 4
clusters.\label{fig:Maximum-a-posteriori-1}}
\end{figure}

The proposed SMM mixture model was applied to the generated data for
various numbers of clusters and the $Q$ function was plotted against
the number of clusters, $|\mathcal{K}|$ as shown in Fig.~\ref{fig:Maximum-a-posteriori-1}.
As can be seen from the figure, the absolute slope of the $Q$ estimates
significantly drops at $|\mathcal{K}|=4$ with estimated $\mathbf{\hat{\pi}}=\{0.169,0.332,0.253,0.246\}$,
which indicates that the true number of clusters and mixture weights
were accurately identified by the SMM estimation model. No similar
clusters were found by the pairwise hypothesis tests discussed in
Sec.~\ref{subsec:Choice-of-clusters}. To assess the accuracy of
the estimated parameters $\hat{\pi}^{(k)},\hat{\mathbf{\rho}}^{(k)},\hat{\mathbf{P}}^{(k)},\hat{\mathbf{H}}^{(k)}$
for each of the estimated clusters, we compared them with the parameters
of the data generating model. The pictorial representation of estimated
and true probabilities is shown in Figures \ref{fig:Data-generating-transition}
and \ref{fig:Computed-clusters'-transition}, respectively. The high
degree of similarity between the plots in these two figures implies
a highly accurate estimation of initial state and transition probabilities.
Additionally, we conducted Chi-square and Kolmogorov-Smirnov tests
to verify the equality of estimated and true parameters. The p-values
of these tests reported in Table \ref{tab:Simulation-and-computed}
are all greater than 0.05, indicating that the equality of estimated
and true parameters (null hypothesis) cannot be rejected, i.e., they
are \emph{statistically} the same at a 95\% confidence level. In summary,
all the results show a clear one-to-one mapping between estimated
and generating (true) cluster parameters, demonstrating the effectiveness
of our SMM clustering model at identifying the underlying parameters
of the patient flow system.

\begin{table}
\protect\protect\protect\caption{p-values for matched cluster parameters. Higher p-values compared
to the significance level indicates that the two compared distributions
were same.\label{tab:Simulation-and-computed}}

\centering{}{\footnotesize{}{}}%
\begin{tabular}{cc|ccc}
{\footnotesize{}{}$i$ }  & {\footnotesize{}{}$j$ }  & {\footnotesize{}{}$\mathbf{\rho}$ }  & {\footnotesize{}{}$\mathbf{P}$ }  & {\footnotesize{}{}$\mathbf{H}$}\tabularnewline
\hline 
{\footnotesize{}{}$1$ }  & {\footnotesize{}{}$2$ }  & {\footnotesize{}{}0.99 }  & {\footnotesize{}{}0.54 }  & {\footnotesize{}{}0.99}\tabularnewline
{\footnotesize{}{}$2$ }  & {\footnotesize{}{}$3$ }  & {\footnotesize{}{}0.99 }  & {\footnotesize{}{}0.17 }  & {\footnotesize{}{}0.83}\tabularnewline
{\footnotesize{}{}$3$ }  & {\footnotesize{}{}$4$ }  & {\footnotesize{}{}0.99 }  & {\footnotesize{}{}0.23 }  & {\footnotesize{}{}0.98}\tabularnewline
{\footnotesize{}{}$4$ }  & {\footnotesize{}{}$1$ }  & {\footnotesize{}{}0.99 }  & {\footnotesize{}{}0.29 }  & {\footnotesize{}{}0.99}\tabularnewline
\hline 
\end{tabular}

\end{table}

Next, we provide deeper insight into the functionality of our SMM method by demonstrating 
the initialization can impact performance and running our method with multiple different 
random initializations helps overcome this challenge. 
To do so we explore three scenarios with 4, 50 and 100 clusters respectively. 
The results are reported in Table~\ref{tab:Clustering-results-under-initializations}.
The table shows the amount of patient data generated in each scenario, 
and the mean and standard deviation of the length of the simulated
patient paths. We remove the data 
where the path is of length 1 (i.e. the patient arrived to the hospital
in a ward for a single time unit and left), since these patients would not be 
considered hospital inpatients. The sample size remaining
is around 75-85\% of the original sample (see column 4).

\begin{table}
\begin{centering}
 \scalebox{0.75}{ \begin{tabular}{>{\raggedright}p{1.9cm}>{\raggedright}p{1.9cm}>{\raggedright}p{2cm}>{\raggedright}p{2cm}|c|cc>{\raggedright}p{1.7cm}}
\textbf{\footnotesize{}\# Clusters, $K$} & \textbf{\footnotesize{}\# Patients, $N$} & \textbf{\footnotesize{}Path lengths' $\mu,\sigma$} & \textbf{\footnotesize{}\# Patients paths with length > 1} & \textbf{\footnotesize{}Run} & \textbf{\footnotesize{}F1-score (\%)} & \textbf{\footnotesize{}Accuracy (\%)} & \textbf{\footnotesize{}Objective value}\tabularnewline
\hline 
\multirow{5}{1.9cm}{{\footnotesize{}4}} & \multirow{5}{1.9cm}{{\footnotesize{}1000}} & \multirow{5}{2cm}{{\footnotesize{}43.2, 32.9}} & \multirow{5}{2cm}{{\footnotesize{}735}} & {\footnotesize{}1} & {\footnotesize{}88.6} & {\footnotesize{}91.6} & {\footnotesize{}208.82}\tabularnewline
 &  &  &  & {\footnotesize{}2} & {\footnotesize{}70.3} & {\footnotesize{}81.4} & {\footnotesize{}201.28}\tabularnewline
 &  &  &  & {\footnotesize{}3} & \textbf{\footnotesize{}88.9} & \textbf{\footnotesize{}91.6} & \textbf{\footnotesize{}209.16}\tabularnewline
 &  &  &  & {\footnotesize{}4} & {\footnotesize{}63.8} & {\footnotesize{}78.2} & {\footnotesize{}199.8}\tabularnewline
 &  &  &  & {\footnotesize{}5} & {\footnotesize{}87.9} & {\footnotesize{}90.3} & {\footnotesize{}208.26}\tabularnewline
\hline 
\multirow{5}{1.9cm}{{\footnotesize{}50}} & \multirow{5}{1.9cm}{{\footnotesize{}5000}} & \multirow{5}{2cm}{{\footnotesize{}47.5, 49.6}} & \multirow{5}{2cm}{{\footnotesize{}4165}} & {\footnotesize{}1} & \textbf{\footnotesize{}90.8} & \textbf{\footnotesize{}92.4} & \textbf{\footnotesize{}620.21}\tabularnewline
 &  &  &  & {\footnotesize{}2} & {\footnotesize{}71.4} & {\footnotesize{}85.7} & {\footnotesize{}601.09}\tabularnewline
 &  &  &  & {\footnotesize{}3} & {\footnotesize{}90.2} & {\footnotesize{}91.9} & {\footnotesize{}619.82}\tabularnewline
 &  &  &  & {\footnotesize{}4} & {\footnotesize{}90.7} & {\footnotesize{}92.4} & {\footnotesize{}620.02}\tabularnewline
 &  &  &  & {\footnotesize{}5} & {\footnotesize{}71.8} & {\footnotesize{}85.1} & {\footnotesize{}601.11}\tabularnewline
\hline 
\multirow{5}{1.9cm}{{\footnotesize{}100}} & \multirow{5}{1.9cm}{{\footnotesize{}10000}} & \multirow{5}{2cm}{{\footnotesize{}51.4, 53.2}} & \multirow{5}{2cm}{{\footnotesize{}8504}} & {\footnotesize{}1} & {\footnotesize{}74.6} & {\footnotesize{}86.7} & {\footnotesize{}1110.57}\tabularnewline
 &  &  &  & {\footnotesize{}2} & {\footnotesize{}74.5} & {\footnotesize{}86.8} & {\footnotesize{}1110.21}\tabularnewline
 &  &  &  & {\footnotesize{}3} & {\footnotesize{}88.4} & {\footnotesize{}90.3} & {\footnotesize{}1121.01}\tabularnewline
 &  &  &  & {\footnotesize{}4} & \textbf{\footnotesize{}89.7} & \textbf{\footnotesize{}91.4} & \textbf{\footnotesize{}1121.89}\tabularnewline
 &  &  &  & {\footnotesize{}5} & {\footnotesize{}74.6} & {\footnotesize{}86.5} & {\footnotesize{}1110.78}\tabularnewline
\hline 
\end{tabular}}
\par\end{centering}
\caption{Clustering results under different initializations.\label{tab:Clustering-results-under-initializations}}

\end{table}

In this table, we show the clustering results from different runs.
Each run has a different random initialization. The number of iterations
in each run was capped at 50. The f1-score and accuracy are shown
as clustering performance measures. The results highlight the differences
in the clustering output for different initializations. As expected,
the value of objective function corresponds to the accuracy levels
\textendash{} higher the objective value, the higher the accuracy. The 
bolded rows of the table indicate the solution that was chosen (out of the 
five random initializations).

With respect to robustness, we found that the estimates of SMM parameters 
in each run (in all three scenarios) were found to
be statistically similar to the true underlying distributions. This shows 
that, while the final cluster assignments may be sensitive to initialization, 
the output of interest, i.e. the semi-Markov parameters, are more robust 
to initialization. As a precaution, however, we suggest 
multiple random initializations as a means to avoid potentially poor 
solutions that may be a result of a particular initialization. 
%As also suggested in Sec.~\ref{subsec:SMM-Clustering-Algorithm},
%similar to most mainstream clustering algorithms, SMM-clustering should
%be performed multiple times with random initializations, and the result
%with the highest objective value should be chosen.

\begin{figure}[h]
\begin{centering}
\includegraphics[scale=0.15]{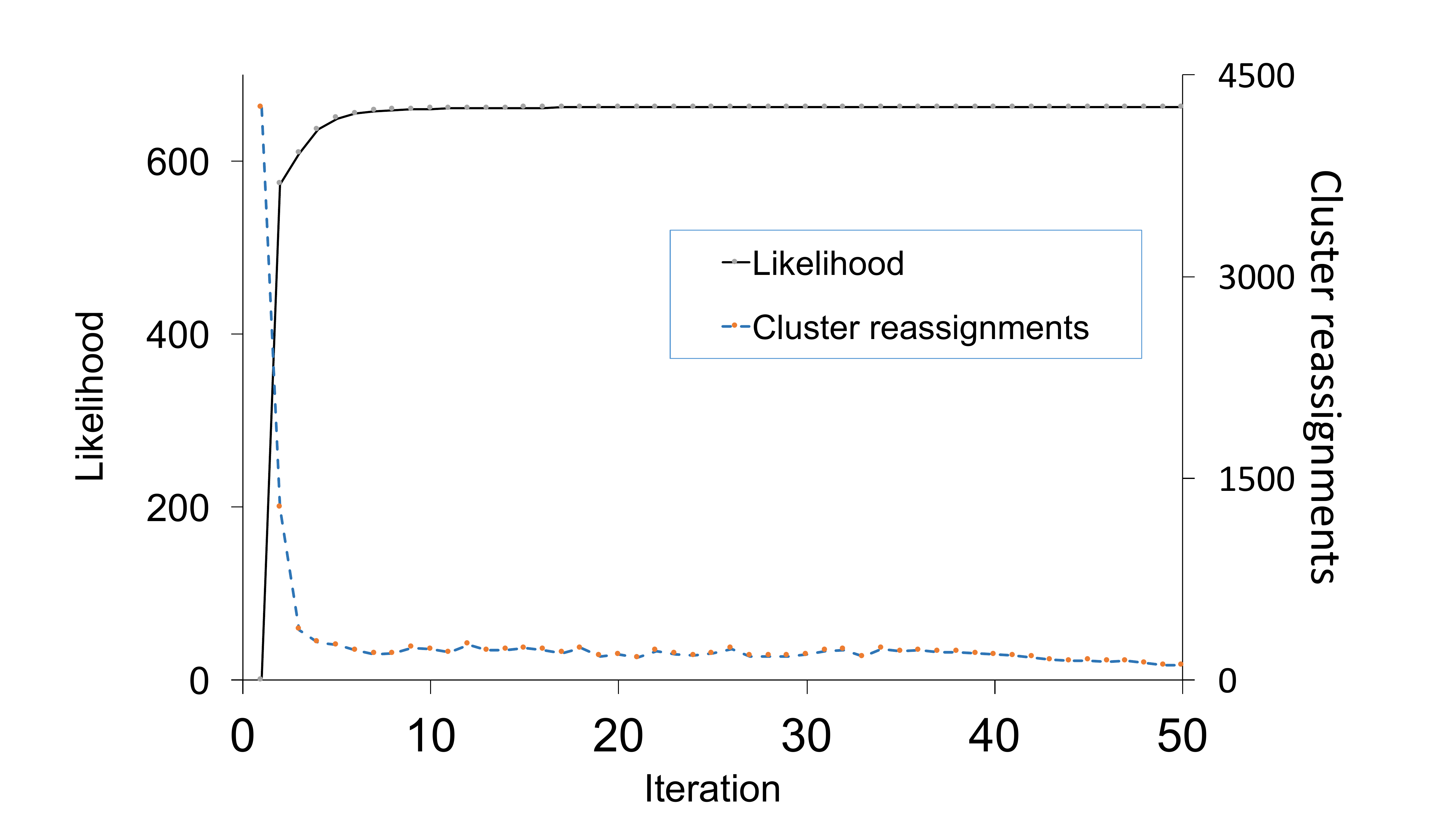}
\par\end{centering}
\caption{Improvement in objective function and reduction in cluster reassignment
as the algorithm convergences.\label{fig:Improvement-in-objective}}
\end{figure}

Additionally, we show the improvement in objective function and the
reduction in cluster reassignment (of trajectories) with each iteration
in Fig.~\ref{fig:Improvement-in-objective}. The figure presents
the result for the problem with 50 clusters. The figure shows that,
(1) the convergence of the algorithm as the iterations progress, and
(2) the objective function reaches a upper bound quickly, but the cluster
assignments keep changing, although very slightly. 
The latter observation indicates that multiple 
cluster solutions gives about the same objective function, though the solution 
quickly becomes relatively stable. This relates
to the identifiability issue and hence our recommendation of setting 
a maximum on the number of iterations.

\vspace{-0.3in}

\subsection{Evaluating the impact of patient flow estimation on scheduling optimization\label{subsec:Evaluating-the-Impact}}

\vspace{-0.2in}

As discussed in Sec.~\ref{subsec:Trajectory-estimation-for} and
\ref{subsec:SMM-Elective-and-Emergency}, the MIP-scheduling model
for resource scheduling (RS) uses the estimated trajectory distribution
of each patient type as an input. In this section, we study the impact
of a better patient path estimation on patient throughput and ward utilization, 
which are the
outcome of RS. To do so, we compare the RS optimization solution using 
CSI with the true optimal (if the actual distribution were known), and 
with four other common patient clustering methods: k-means clustering, 
DRG-based clustering, Gaussian clustering, and Markov model-based clustering. 

We compare our CSI method with the commonly used attribute-based clustering 
methods: Diagnosis Related Group (DRG)-based clustering and k-means attribute 
clustering. In these methods, patients are first
clustered into groups based on the similarity of 
their personal and medical attributes. 
While DRG-based clustering uses the patients' diagnosis (disease) for grouping,
k-means uses other attributes like age, sex, etc. for a finer grouping.
The path distribution for each patient cluster are then estimated empirically.
Details are given in Online Appendix-C.

There can be drawbacks to k-means clustering. Sometimes 
the data density can be non-convex or severely unbalanced, which affects 
the method's effectiveness. Hence, we also compare our method 
with a Gaussian mixture distribution approach for clustering the patients
based on their attributes. We also implement and compare our method with 
a mixture Markov model based clustering method (Cadez et al., 2000). 
This Markov model clustering is different from SMM due to its assumption 
that the holding time distribution is independent of state transition.

The patient path distributions, computed from the patient clusters,
are supplied to the MIP model (the \emph{maximum elective admission}
formulation presented in Sec.~\ref{subsec:Mixed-Integer-Programming}
and Online Appendix B). In Table~\ref{tab:Simulation-Service-level-improvements},
 we present the optimization results for the objectives of maximizing throughput 
and ward utilization for the same three scenarios studied in the previous 
section; cluster sizes $K\in\{4,50,100\}$.

\begin{table}[h]
\begin{centering}
\scalebox{0.8}{
\begin{tabular}{l|>{\raggedright}p{1.9cm}>{\raggedright}p{1.7cm}|>{\raggedright}p{1.9cm}>{\raggedright}p{1.7cm}|>{\raggedright}p{1.9cm}>{\raggedright}p{1.7cm}}
 & \multicolumn{2}{c|}{{\footnotesize{}$K=4$}} & \multicolumn{2}{c|}{{\footnotesize{}$K=50$}} & \multicolumn{2}{c}{{\footnotesize{}$K=100$}}\tabularnewline
\cline{2-7} 
 & \textbf{\footnotesize{}Throughput } & \textbf{\footnotesize{}Utilization} & \textbf{\footnotesize{}Throughput } & \textbf{\footnotesize{}Utilization} & \textbf{\footnotesize{}Throughput} & \textbf{\footnotesize{}Utilization}\tabularnewline
\hline 
{\footnotesize{}(Optimal)} & \textbf{\footnotesize{}85\%} & \textbf{\footnotesize{}49\%} & \textbf{\footnotesize{}78\%} & \textbf{\footnotesize{}34\%} & \textbf{\footnotesize{}89\%} & \textbf{\footnotesize{}45\%}\tabularnewline
{\footnotesize{}CSI} & \textbf{\footnotesize{}81\%} & \textbf{\footnotesize{}49\%} & \textbf{\footnotesize{}75\%} & \textbf{\footnotesize{}33\%} & \textbf{\footnotesize{}88\%} & \textbf{\footnotesize{}45\%}\tabularnewline
{\footnotesize{}DRG} & {\footnotesize{}21\%} & {\footnotesize{}11\%} & {\footnotesize{}24\%} & {\footnotesize{}14\%} & {\footnotesize{}28\%} & {\footnotesize{}19\%}\tabularnewline
{\footnotesize{}k-means} & {\footnotesize{}24\%} & {\footnotesize{}24\%} & {\footnotesize{}25\%} & {\footnotesize{}16\%} & {\footnotesize{}32\%} & {\footnotesize{}21\%}\tabularnewline
{\footnotesize{}Gaussian} & {\footnotesize{}19\%} & {\footnotesize{}9\%} & {\footnotesize{}18\%} & {\footnotesize{}10\%} & {\footnotesize{}21\%} & {\footnotesize{}10\%}\tabularnewline
{\footnotesize{}Markov} & {\footnotesize{}58\%} & {\footnotesize{}38\%} & {\footnotesize{}51\%} & {\footnotesize{}21\%} & {\footnotesize{}61\%} & {\footnotesize{}37\%}\tabularnewline
\hline 
\end{tabular}}
\par\end{centering}
\caption{Percentage increase in service level metrics: throughput from the
number of elective patient admission, and ward utilization.\label{tab:Simulation-Service-level-improvements}}
\end{table}

The table contains a row for ``optimal''. Here the ``optimal''
result is drawn by using the known true underlying number of patient
types and their path distributions. This result serves as the baseline
(or, the upper bound in this case) to assess the performance of other
methods. 

As shown in Table~\ref{tab:Simulation-Service-level-improvements},
the outcome from CSI is quite close to the optimal. The Markov model 
is the next best method, but falls well short of CSI. 
This is because, similar to the SMM-clustering used in CSI, 
the Markov model clustering also groups 
the patients based on similarity in their paths. However, the performance is 
worse than the one from SMM because it assumes a same holding time distribution, i.e. the distribution
on length-of-stay in a ward (before moving to another)
is always the same. This means that the Markov model is ignoring a critical 
feature of ward interactions, i.e. that holding time and ward transitions 
are dependent.

The scheduling outcome from empirical patient path distributions (derivation
expressions in Online Appendix-D) drawn from attribute-based clustering,
viz. k-means, DRG and Gaussian, were significantly poorer than the
optimal. Among them, Gaussian performed the worst because of its ineffectiveness
in clustering categorical variables present in patient attributes
data.

\subsection{Applying SMM-clustering in practice}
In this section, we discuss some of the advantages and disadvantages of the 
SMM clustering method and what types of problems are best suited for 
applying our method.

% Our SMM clustering approach provides accurate SMM parameter
% estimates even if the cluster assignment of all the training data is not
% the global optimal. For the patient clustering problem for hospital
% resource optimization, the SMM-clustering approach effectively captures
% the interactions between the hospital wards. Since, it models the
% patient paths from their observed movements between wards, the underlying
% relationships between the wards are drawn from the data. This is important because these patients
% will require different levels of hospital resources.

The results of this simulation study show that the proposed CSI yields
a schedule in RS that is very near the true optimal, and significantly
outperforms existing HASC methods. Our SMM clustering model outperforms traditional 
attribute-based clustering methods 
specifically because it takes trajectories into account in the clustering process. 
Attribute-based clustering, on the other hand, relies on an indirect relationship between 
attributes and patient trajectories rather than directly employing trajectories as 
a clustering approach. This highlights one of the key innovations of our new method. 
Instructively, our SMM-based clustering method also significantly outperforms the Markov 
model-based clustering method, which \emph{does} consider patient trajectories. This highlights 
a second innovation of our SMM method, which is that we properly consider the interaction 
between wards. That is, our method allows the holding time distribution and ward transitions 
to be dependent, which is ignored in the Markov model-based clustering method. 
In addition, SMM clustering effectively differentiates between patients with different
lengths-of-stay at the wards. This more subtle modeling difference turns out to have a 
significant impact on model performance.

As a result, we find that our SMM-clustering approach is most effective being applied to 
problems with the following characteristics. First, our method performs well in situations 
where individual characteristics available in the data (e.g. age, sex, co-morbidity) are 
not adequately explanatory of trajectories. From a patient flow perspective, this feature 
is highlighted by the example described by Fig. \ref{fig:Case-Study-Trajectories-of-patients}. 
Clearly there are applications outside of patient flow 
that share this feature. 

Second, problems with complex and dependent network interactions can cause simpler methods 
to perform poorly, creating significant opportunity for our method 
to outperform existing clustering methods. In particular, non-Markovian networks 
benefit significantly from relaxing the Markovian assumption in the clustering 
methodology, as ours does. Non-Markovian networks are quite common in healthcare, 
since a patient's history has been shown in many contexts to correlate with 
future requirements and outcomes. However, this feature is not unique to patient flow 
systems. 

Finally, in our context the 
output of interest is not the clusters themselves, but rather the 
trajectory distributions derived from the clusters. We have shown that our 
model is quite robust to local optima and intializations in terms of 
the overall trajectory distributions because near the optimal EM solution,  
clusters may continue to change but the overall distributions derived 
from each cluster remain relatively stable. Our SMM clustering 
approach provides accurate parameter estimates even if the cluster 
assignment of all the training data is not the global optimal. This also helps mitigate 
identifiability concerns, since different many clustering solutions can 
generate very similar holding time and transition distributions, which is 
the output of interest. Thus applications that 
rely on the semi-Markov distributional output rather than the 
actual clustering results themselves are best suited for SMM.

SMM-clustering, however, does rely on Markovian transitions between 
wards. While this is often not strictly true in patient flow 
systems, much past literature has found this modeling assumption to 
be sufficiently accurate. Performance may suffer, however, if transitions 
are strongly history dependent, 
thereby making the Markovian assumption a poor representation of reality. 
In such a situation, the trajectory data can be tested for 
different orders of the Markovian property and the state space may be 
able to be expanded to restore the Markovian property if necessary. 

Another disadvantage is that, due to athe large number
of parameters to estimate, SMM-clustering requires large amount of
data. While, large amounts of data and an underlying Markovian property
is common for patient flow problems, other methods such
as in Ranjan et. al (2016) can be incorporated to mitigate data scarcity issues.

%This study is the first, to our knowledge, to quantify the impact
%of estimation approach on elective inpatient optimization solutions,
%and further demonstrates the importance of effective estimation techniques
%(e.g. our SMM method) on patient flow optimization.

\vspace{-0.2in}

\section{Case study on real hospital data\label{sec:Case-study-on}}

\vspace{-0.2in}

In this section, we will study the impact of our integrated framework
(CSI) on hospital resource optimization at a partner hospital, and
as a holistic tool for the HASC problem. In particular, we focus on validating
the trajectory estimation and RS models, as forecasting arrival streams
is out of the scope of this paper. Hence, we take the arrival stream
as given in order to independently evaluate the accuracy and impact
of trajectory estimation on the HASC problem. %As explained in methodology overview%(Section \ref{fig:Methodology-overview}), the developed CM module%can be used for census forecasting as well. However, since %the arrival rate of elective patients is deterministic, the census forecasting primarily depends on the accurate trajectory estimation. Hence, we%will focus on demonstrating the performance of our approach for accurate%trajectory estimation and RS. 

We use historical data of patient admission and transitions in a hospital
with 55 wards including surgical, ICU/CCU, medicine, neurology, oncology,
obstetrics, etc. Although, physically the hospital has more than 55
wards, for simplicity several wards were grouped based on expert prior
knowledge about their similarity. This system is a good example of
a complex hospital system with general ward network structure, transfers
and blocking/congestion.

We obtained one year of data from 2012, with about 11,000 patients
who stayed at least one night in the hospital. The data set includes
the patient flow data, length-of-stay at each ward, and patient attribute
data, for e.g. age, sex, diagnosis, etc. The ratio of elective and
emergency patients in the data is almost equal. Patients have an average
of 4.1 transfers before leaving the hospital.
We compare the performance of the CSI model with that of the established 
clustering and estimation approaches.

\begin{figure}[h]
\begin{centering}
\includegraphics[scale=0.5]{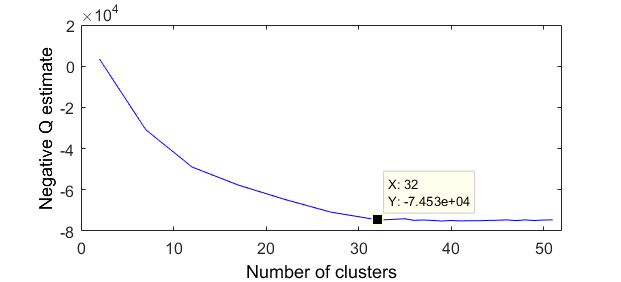} 
\par\end{centering}
\caption{The estimated Q function against increasing number of clusters for
the real data in Case Study. It is observed that the improvement in
Q function is not significant after 32 clusters. \label{fig:Case-Study-Q-function}}
\end{figure}

We begin with the CM step by applying SMM-based clustering on patient
trajectory data to identify patient types. From Fig.~\ref{fig:Case-Study-Q-function},
we can infer that there are 32 patient types. Again, no redundant
clusters were found from pairwise hypothesis testing. The trajectory
probability distributions for each of these patient types are computed
using Eq.~\ref{eq:gamma-input to schedule}. Simultaneously conventional
partition-based clustering methods, discussed above in 
Sec.~\ref{subsec:Evaluating-the-Impact},
viz. $k$-means, DRG and Gaussian clustering, are used to cluster patients
based on the patient's attribute data. 

While for DRG, the number of clusters is found from the data (the number
of diagnosis types), the criteria for finding the optimal number of
clusters with $k$-means and Gaussian are rather subjective. Therefore,
in order to have a fair comparison, we use the same number of clusters
as chosen by SMM (i.e., 32 clusters). This does not affect
the optimization in RS even if we have a few redundant clusters, but
prevents the risk of suboptimal results due to under-estimation of
the number of clusters. Therefore, the benefits demonstrated by this
case study represent a conservative estimate of the true potential
benefits when compared to an application to a hospital in the real
world. After performing the attribute-based clustering, empirical
trajectory distributions are estimated for each patient cluster 
using the same approach regardless of clustering method.

\begin{figure}[ht]
\begin{centering}
\begin{minipage}[t]{0.46\columnwidth}%
\begin{center}
\subfloat[Patient trajectories from a $k$-means cluster\label{fig:Case-Study-naive-trajectories}]{\begin{centering}
\includegraphics[scale=0.5]{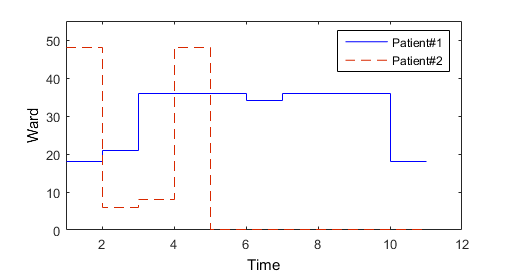} 
\par\end{centering}
\centering{}

}
\par\end{center}%
\end{minipage}%
\begin{minipage}[t]{0.45\columnwidth}%
\subfloat[Patient trajectories from a SMM based cluster\label{fig:Case-Study-SMM-trajectories}]{\begin{centering}
\includegraphics[scale=0.5]{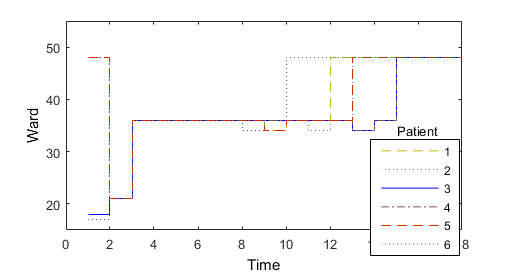} 
\par\end{centering}
\centering{}

}%
\end{minipage}
\par\end{centering}
\caption{Trajectories of patients belonging to same cluster. It is observed
that patients in SMM based clusters follow more similar trajectories
than $k$-means. \label{fig:Case-Study-Trajectories-of-patients}}
\end{figure}

To verify our claims that two patients with similar attributes may
not follow the same trajectory, we observed two patients who were
put into the same cluster using the $k$-means; they were both male,
aged between 55-65 years and were diagnosed for heart disease. Their
trajectories within hospital are shown in Fig.~\ref{fig:Case-Study-naive-trajectories}.
In this figure, patient\#1 enters the cardiology ward, transitions
to the angiography center then to the neurology ward and finally back
to cardiology before leaving the hospital. Patient\#2, on the other
hand, begins their stay in the surgical ward, transitions to the heart
clinic, then the ICU, then the operating theater, then to the ICU
again and finally back to the surgical ward before being discharged.
Although the observed attributes for both patients show similar profiles
and a heart disease diagnosis, the trajectories followed by these
patients were very different. Observing their trajectories more closely,
one can see that patient\#2 might have had a severe heart condition,
while patient\#1 had a relatively milder heart condition only requiring
angiography.

When employing our SMM-clustering method, we do not see such dissimilarity
in patient trajectories within one cluster. As an example, Fig.~\ref{fig:Case-Study-SMM-trajectories}
shows trajectories of a few patients from one of the clusters identified
by the SMM approach. Most of the patients in this cluster enter the
hospital either in surgical or cardiology wards, then transition to
the heart clinic, ICU, operating theater and finally cardiology before
leaving the hospital. There is one case of patient\#6 who entered
the hospital in ortho and spine center, but then followed similar
trajectory of going to heart clinic, ICU, operating theater and finally
cardiology. This could be caused by a heart condition developing during
an orthopedic admission, or possibly due to initial off-ward placement
(because the cardiology ward was full). It is interesting to see that
if we would have used the conventional attribute based clustering
this patient would have been put into a orthopedic related cluster,
while the SMM approach was able to identify the patient's ``true''
cluster. %based on the complete trajectory. Thus, SMM-clustering provides a higher chance of finding the ``true'' patient type compared with the traditional, attribute-based clustering approach.

To test the impact of our SMM approach on the RS optimization, 
we use the \emph{maximum
elective admission} formulation given in Sec.~\ref{subsec:Mixed-Integer-Programming}
and Online Appendix B. The goal is to increase the volume of patients
served (throughput), thereby increasing revenues, while maintaining
the same level of service and access. The results are shown in Table~\ref{tab:Case-Study-Comparing-the-improvement}
for the CSI and other methods relative to the baseline current elective
admission schedule of the partner hospital. 

Our CSI method demonstrates a potential increase in elective admissions
of 97\%, and an increase of 22\% for ward utilization. Similar to
the simulation study, the Markov clustering performs second best, 
with improvements of 63\% and 19\% for throughput and
utilization, respectively. Among the other methods, k-means performs 
the best, yet significantly worse than CSI, with improvement of 30\%
and 8\%. In practice, attribute partition methods like k-means and
DRG are commonly used, though they leave much to be desired.

This case study of a partner hospital demonstrates the importance
of an accurate patient clustering and trajectory estimation method,
as using our CSI not only provides a more accurate forecast of the
hospital stochastic workload process, but also dramatically improves
optimization solutions. Further, to the best of our knowledge, our
CSI method is the only approach in the extant literature that has
all the properties required for effective integration with admission
scheduling optimization approaches: \emph{scalable} to hospital of
any size, considers \emph{ward interactions}, and accounts for \emph{patient
heterogeneity}.

\begin{table}
\begin{centering}
\begin{tabular}{l|cc}
 & \textbf{\footnotesize{}Throughput} & \textbf{\footnotesize{}Utilization}\tabularnewline
\hline 
{\footnotesize{}CSI} & \textbf{\footnotesize{}97\%} & \textbf{\footnotesize{}22\%}\tabularnewline
{\footnotesize{}DRG} & {\footnotesize{}26\%} & {\footnotesize{}11\%}\tabularnewline
{\footnotesize{}k-means} & {\footnotesize{}30\%} & {\footnotesize{}8\%}\tabularnewline
{\footnotesize{}Gaussian} & {\footnotesize{}22\%} & {\footnotesize{}5\%}\tabularnewline
{\footnotesize{}Markov} & {\footnotesize{}63\%} & {\footnotesize{}19\%}\tabularnewline
\hline 
\end{tabular}
\par\end{centering}
\caption{Comparing the percentage improvement in throughput (elective admissions)
and ward utilization (workload) for proposed CSI and other traditional
methods with respect to the current service and workload levels using
real hospital data.\label{tab:Case-Study-Comparing-the-improvement}}
\end{table}

\vspace{-0.2in}

\section{Conclusion \label{sec:Conclusion-and-future}}

\vspace{-0.2in}

The \emph{Hospital Admission Scheduling and Control} problem is comprised
of two main components: \emph{census modeling} and \emph{resource
scheduling}. Previous work on this long-standing problem has considered
one or the other, but not both. In this paper we develop a new method
based on \emph{semi-Markov model} (SMM) based clustering for identifying
patient type clusters and estimating cluster trajectory distributions
that integrates seamlessly with existing scheduling optimization approaches.
This integration is proven to be extremely important, as optimal solutions
using our SMM approach dramatically outperform optimal solutions using
the traditional empirical estimation techniques.

As a theoretical contribution, our novel approach is able to model
an entire hospital of any size as a coordinated system with a complex,
general network of wards and patient transitions between them. Further,
the model has been shown to be \emph{scalable}, accounts for \emph{ward
interactions}, and for patient \emph{heterogeneity}, which has not
been previously achieved by other methods in the literature. Further,
our SMM-clustering is a general purpose algorithm applicable to any
movement or sequence data having spatial and temporal dimension, for
example, \emph{clickstream} data of users on a website or movement
of cell-phone users among a network of towers.

Our SMM approach was designed to integrate with RS approaches that
provide an optimal controllable schedule by patient type for each
day of week so this approach can be adopted by any specialty or multi-specialty
hospital for streamlining their procedures, stabilizing the operating
environment for their personnel, improving utilization of hospital
resources, and enabling cost savings for both patients and hospitals. The automated,
algorithmic approach to clustering and trajectory estimation is also
appealing compared to ad-hoc, manual, and heuristic approaches currently
employed in practice (which can take months to implement and are difficult
to validate statistically).

The SMM-clustering method was validated by simulating data from \emph{known}
generating mixture distributions. The SMM estimated clusters and their
distributions were found to be statistically the same as the generating
mixture distributions at a 95\% confidence level. 
%We validated the
%efficacy of the CSI method under simulated conditions based on a small
%hospital scenario, where we know the true number of patient types,
%how they are clustered, and their trajectory distributions. 
Optimizing
the elective schedule based on inputs from our SMM method achieved
outcomes that were very close to the the ``true optimum'' (i.e.
given perfect knowledge of patient flow dynamics) while the existing
traditional method gave performed significantly worse.

A case study using real hospital data showed that the number of elective
admissions could be increased by 97\% (with the same level of access)
compared to only a 30\% increase using traditional empirical methods
(which are comparable to previous optimization improvements reported
in the literature). Moreover, the average ward utilization could be
improved by 22\% using our approach compared with only an 8\% improvement
using the traditional approach.

In conclusion, our approach develops a novel method for spatio-temporal
clustering and trajectory estimation that has a profound impact on an important
patient flow problem with the potential to improve revenues and/or
cost, quality, and access to care.

\bibliographystyle{apa}
\nocite{*}
\begin{spacing}{0.9}
\bibliography{SMM-poms-biblio}

\begin{thebibliography}{}

\bibitem[\protect\astroncite{Abraham et~al.}{2009}]{abraham2009short}
Abraham, G., Byrnes, G.~B., and Bain, C.~A. (2009).
\newblock Short-term forecasting of emergency inpatient flow.
\newblock {\em IEEE Transactions on Information Technology in Biomedicine},
  13(3):380--388.

\bibitem[\protect\astroncite{Adan et~al.}{2009}]{adan2009patient}
Adan, I., Bekkers, J., Dellaert, N., Vissers, J., and Yu, X. (2009).
\newblock Patient mix optimisation and stochastic resource requirements: A case
  study in cardiothoracic surgery planning.
\newblock {\em Health care management science}, 12(2):129--141.

\bibitem[\protect\astroncite{Aiken et~al.}{2002}]{aiken2002hospital}
Aiken, L.~H., Clarke, S.~P., Sloane, D.~M., Sochalski, J., and Silber, J.~H.
  (2002).
\newblock Hospital nurse staffing and patient mortality, nurse burnout, and job
  dissatisfaction.
\newblock {\em Jama}, 288(16):1987--1993.

\bibitem[\protect\astroncite{Armony et~al.}{2015}]{armony2015patient}
Armony, M., Israelit, S., Mandelbaum, A., Marmor, Y.~N., Tseytlin, Y., and
  Yom-Tov, G.~B. (2015).
\newblock On patient flow in hospitals: A data-based queueing-science
  perspective. an extended version (ev).
\newblock Technical report, Working paper, http://ie. technion. ac.
  il/serveng/References/Patient\% 20flow\% 20main. pdf.

\bibitem[\protect\astroncite{Bekker and Koeleman}{2011}]{bekker2011scheduling}
Bekker, R. and Koeleman, P.~M. (2011).
\newblock Scheduling admissions and reducing variability in bed demand.
\newblock {\em Health care management science}, 14(3):237--249.

\bibitem[\protect\astroncite{Billingsley}{1960}]{billingsley1960statistical}
Billingsley, P. (1960).
\newblock Statistical inference for markov processes.

\bibitem[\protect\astroncite{Billingsley}{1961}]{billingsley1961statistical}
Billingsley, P. (1961).
\newblock Statistical methods in markov chains.
\newblock {\em The Annals of Mathematical Statistics}, pages 12--40.

\bibitem[\protect\astroncite{Cadez et~al.}{2000}]{cadez2000visualization}
Cadez, I., Heckerman, D., Meek, C., Smyth, P., and White, S. (2000).
\newblock Visualization of navigation patterns on a web site using model-based
  clustering.
\newblock In {\em Proceedings of the sixth ACM SIGKDD international conference
  on Knowledge discovery and data mining}, pages 280--284. ACM.

\bibitem[\protect\astroncite{Earnest et~al.}{2005}]{earnest2005using}
Earnest, A., Chen, M.~I., Ng, D., and Sin, L.~Y. (2005).
\newblock Using autoregressive integrated moving average (arima) models to
  predict and monitor the number of beds occupied during a sars outbreak in a
  tertiary hospital in singapore.
\newblock {\em BMC Health Services Research}, 5(1):1.

\bibitem[\protect\astroncite{Faddy and McClean}{1999}]{faddy1999analysing}
Faddy, M. and McClean, S. (1999).
\newblock Analysing data on lengths of stay of hospital patients using
  phase-type distributions.
\newblock {\em Applied Stochastic Models in Business and Industry},
  15(4):311--317.

\bibitem[\protect\astroncite{Fetter et~al.}{1980}]{fetter1980case}
Fetter, R.~B., Shin, Y., Freeman, J.~L., Averill, R.~F., and Thompson, J.~D.
  (1980).
\newblock Case mix definition by diagnosis-related groups.
\newblock {\em Medical care}, 18(2):i--53.

\bibitem[\protect\astroncite{Green}{2006}]{green2006queueing}
Green, L. (2006).
\newblock Queueing analysis in healthcare.
\newblock In {\em Patient flow: reducing delay in healthcare delivery}, pages
  281--307. Springer.

\bibitem[\protect\astroncite{Griffin et~al.}{2012}]{griffin2012improving}
Griffin, J., Xia, S., Peng, S., and Keskinocak, P. (2012).
\newblock Improving patient flow in an obstetric unit.
\newblock {\em Health care management science}, 15(1):1--14.

\bibitem[\protect\astroncite{Griffith et~al.}{1976}]{griffith1976cost}
Griffith, J.~R., Hancock, W.~M., and Munson, F.~C. (1976).
\newblock {\em Cost control in hospitals}.
\newblock Health Administration Press.

\bibitem[\protect\astroncite{Hall et~al.}{2006}]{hall2006modeling}
Hall, R., Belson, D., Murali, P., and Dessouky, M. (2006).
\newblock Modeling patient flows through the healthcare system.
\newblock In {\em Patient flow: Reducing delay in healthcare delivery}, pages
  1--44. Springer.

\bibitem[\protect\astroncite{Hancock and Walter}{1979}]{hancock1979use}
Hancock, W.~M. and Walter, P.~F. (1979).
\newblock The use of computer simulation to develop hospital systems.
\newblock {\em ACM SIGSIM Simulation Digest}, 10(4):28--32.

\bibitem[\protect\astroncite{Hancock and Walter}{1983}]{hancock1983ascs}
Hancock, W.~M. and Walter, P.~F. (1983).
\newblock {\em The" ASCS": Inpatient Admission Scheduling and Control System}.
\newblock Health Administration Press.

\bibitem[\protect\astroncite{Harper}{2005}]{harper2005review}
Harper, P.~R. (2005).
\newblock A review and comparison of classification algorithms for medical
  decision making.
\newblock {\em Health Policy}, 71(3):315--331.

\bibitem[\protect\astroncite{Harper and Shahani}{2002}]{harper2002modelling}
Harper, P.~R. and Shahani, A. (2002).
\newblock Modelling for the planning and management of bed capacities in
  hospitals.
\newblock {\em Journal of the Operational Research Society}, 53(1):11--18.

\bibitem[\protect\astroncite{Helm and Van~Oyen}{2014}]{helm2014design}
Helm, J.~E. and Van~Oyen, M.~P. (2014).
\newblock Design and optimization methods for elective hospital admissions.
\newblock {\em Operations Research}, 62(6):1265--1282.

\bibitem[\protect\astroncite{Irvine et~al.}{1994}]{irvine1994stochastic}
Irvine, V., McClean, S., and Millard, P. (1994).
\newblock Stochastic models for geriatric in-patient behaviour.
\newblock {\em Mathematical Medicine and Biology}, 11(3):207--216.

\bibitem[\protect\astroncite{Jacobson et~al.}{2006}]{jacobson2006discrete}
Jacobson, S.~H., Hall, S.~N., and Swisher, J.~R. (2006).
\newblock Discrete-event simulation of health care systems.
\newblock In {\em Patient flow: Reducing delay in healthcare delivery}, pages
  211--252. Springer.

\bibitem[\protect\astroncite{Jones et~al.}{2002}]{jones2002forecasting}
Jones, S.~A., Joy, M.~P., and Pearson, J. (2002).
\newblock Forecasting demand of emergency care.
\newblock {\em Health care management science}, 5(4):297--305.

\bibitem[\protect\astroncite{Kao}{1972}]{kao1972semi}
Kao, E.~P. (1972).
\newblock A semi-markov model to predict recovery progress of coronary
  patients.
\newblock {\em Health Services Research}, 7(3):191.

\bibitem[\protect\astroncite{Kao}{1974}]{kao1974modeling}
Kao, E.~P. (1974).
\newblock Modeling the movement of coronary patients within a hospital by
  semi-markov processes.
\newblock {\em Operations Research}, 22(4):683--699.

\bibitem[\protect\astroncite{Keehan et~al.}{2007}]{keehan2007expenses}
Keehan, S., Sisko, A., and Truffer, C. (2007).
\newblock Expenses for hospital inpatient stays: 2004.
\newblock {\em Statistical Brief}, 164.

\bibitem[\protect\astroncite{Konrad et~al.}{2013}]{konrad2013modeling}
Konrad, R., DeSotto, K., Grocela, A., McAuley, P., Wang, J., Lyons, J., and
  Bruin, M. (2013).
\newblock Modeling the impact of changing patient flow processes in an
  emergency department: Insights from a computer simulation study.
\newblock {\em Operations Research for Health Care}, 2(4):66--74.

\bibitem[\protect\astroncite{Littig and Isken}{2007}]{littig2007short}
Littig, S.~J. and Isken, M.~W. (2007).
\newblock Short term hospital occupancy prediction.
\newblock {\em Health care management science}, 10(1):47--66.

\bibitem[\protect\astroncite{Marshall and
  McClean}{2003}]{marshall2003conditional}
Marshall, A. and McClean, S. (2003).
\newblock Conditional phase-type distributions for modelling patient length of
  stay in hospital.
\newblock {\em International Transactions in Operational Research},
  10(6):565--576.

\bibitem[\protect\astroncite{Massey and Whitt}{1993}]{massey1993networks}
Massey, W.~A. and Whitt, W. (1993).
\newblock Networks of infinite-server queues with nonstationary poisson input.
\newblock {\em Queueing Systems}, 13(1-3):183--250.

\bibitem[\protect\astroncite{McLachlan and
  Krishnan}{2007}]{mclachlan2007algorithm}
McLachlan, G. and Krishnan, T. (2007).
\newblock {\em The EM algorithm and extensions}, volume 382.
\newblock John Wiley \& Sons.

\bibitem[\protect\astroncite{Ranjan et~al.}{2016}]{ranjan2016sequence}
Ranjan, C., Ebrahimi, S., and Paynabar, K. (2016).
\newblock Sequence graph transform (sgt): A feature extraction function for
  sequence data mining.
\newblock {\em arXiv preprint arXiv:1608.03533}.

\bibitem[\protect\astroncite{Richardson et~al.}{2006}]{richardson2006increase}
Richardson, D.~B. et~al. (2006).
\newblock Increase in patient mortality at 10 days associated with emergency
  department overcrowding.
\newblock {\em Medical journal of Australia}, 184(5):213.

\bibitem[\protect\astroncite{Ridley et~al.}{1998}]{ridley1998classification}
Ridley, S., Jones, S., Shahani, A., Brampton, W., Nielsen, M., and Rowan, K.
  (1998).
\newblock Classification treesa possible method for iso-resource grouping in
  intensive care.
\newblock {\em Anaesthesia}, 53(9):833--840.

\bibitem[\protect\astroncite{Smallwood et~al.}{1969}]{smallwood1969medical}
Smallwood, R., Murray, G., Silva, D., Sondik, E., and Klainer, L. (1969).
\newblock A medical service requirements model for health system design.
\newblock {\em Proceedings of the IEEE}, 57(11):1880--1887.

\bibitem[\protect\astroncite{Taylor et~al.}{2000}]{taylor2000stochastic}
Taylor, G., McClean, S., and Millard, P. (2000).
\newblock Stochastic models of geriatric patient bed occupancy behaviour.
\newblock {\em Journal of the Royal Statistical Society: Series A (Statistics
  in Society)}, 163(1):39--48.

\bibitem[\protect\astroncite{Thomas}{1968}]{thomas1968model}
Thomas, W.~H. (1968).
\newblock A model for predicting recovery progress of coronary patients.
\newblock {\em Health Services Research}, 3(3):185.

\bibitem[\protect\astroncite{Weiss et~al.}{1982}]{weiss1982iterative}
Weiss, E.~N., Cohen, M.~A., and Hershey, J.~C. (1982).
\newblock An iterative estimation and validation procedure for specification of
  semi-markov models with application to hospital patient flow.
\newblock {\em Operations Research}, 30(6):1082--1104.

\bibitem[\protect\astroncite{Wu}{1983}]{wu1983convergence}
Wu, C.~J. (1983).
\newblock On the convergence properties of the em algorithm.
\newblock {\em The Annals of statistics}, pages 95--103.

\bibitem[\protect\astroncite{Zeltyn et~al.}{2011}]{zeltyn2011simulation}
Zeltyn, S., Marmor, Y.~N., Mandelbaum, A., Carmeli, B., Greenshpan, O., Mesika,
  Y., Wasserkrug, S., Vortman, P., Shtub, A., Lauterman, T., et~al. (2011).
\newblock Simulation-based models of emergency departments:: Operational,
  tactical, and strategic staffing.
\newblock {\em ACM Transactions on Modeling and Computer Simulation (TOMACS)},
  21(4):24.

\bibitem[\protect\astroncite{Zhang et~al.}{2009}]{zhang2009mixed}
Zhang, B., Murali, P., Dessouky, M., and Belson, D. (2009).
\newblock A mixed integer programming approach for allocating operating room
  capacity.
\newblock {\em Journal of the Operational Research Society}, 60(5):663--673.

\end{thebibliography}
\end{spacing}
\pagebreak{}
\begin{center}
{\LARGE{}The Impact of Estimation: A New Method for Clustering and
Trajectory Estimation in Patient Flow Modeling}
\par\end{center}{\LARGE \par}

\section*{Appendices}

\subsection*{Appendix A: Derivation of SMM-clustering update expressions for EM
algorithm\label{app:Appendix-A:-Derivation}}

In this appendix, we present the derivation of parameter update expressions
for the EM algorithm in Sec.~\ref{subsec:EM-Algorithm}. As mentioned
in the section, we have to obtain the posterior distributions of the
parameters to find their optimal estimates that maximizes Eq.~\ref{eq:Q function def for semi markov}.

We use Dirichlet prior distributions, given in Eq.~\ref{eq:dirichlet},
for the parameters. The Dirichlet hyperparameters for parameters in
$\mathbf{\Theta}=\{\pi^{(k)},\pmb{\rho}^{(k)},\mathbf{P}^{(k)},\mathbf{H}^{(k)}\},k\in\mathcal{K}$
are denoted by $\{a_{\pi}^{(k)},a_{\rho}^{(k)},a_{P}^{(k)},a_{H}^{(k)}\},k\in\mathcal{K}$,
respectively. For each model parameter, the hyperparameters can be
set to equal values, if there is no specific prior knowledge (non-informative
prior). Besides, we assume the parameters are independent. Using it
with the conditions on probability sums equal to 1 in Eq.~\ref{eq:Prob-sum-1}
and parameter independence assumptions gives the following expressions
for prior probabilities,

\begin{eqnarray}
p(\pi) & \propto & \prod_{k\in\mathcal{K}}\left(\pi^{(k)}\right)^{a_{\pi}^{(k)}-1}\nonumber \\
p(\pmb{\rho}) & \propto & \prod_{k\in\mathcal{K}}\prod_{u\in\mathcal{U}}\left(\rho_{u}^{(k)}\right)^{a_{\rho,u}^{(k)}-1}\nonumber \\
p(\mathbb{\mathbf{P}}) & \propto & \prod_{k\in\mathcal{K}}\prod_{u\in\mathcal{U}}\prod_{j\in\mathcal{U}}\left(P_{uj}^{(k)}\right)^{a_{P,uj}^{(k)}-1}\nonumber \\
p(\mathbf{H}) & \propto & \prod_{k\in\mathcal{K}}\prod_{u\in\mathcal{U}}\prod_{j\in\mathcal{U}}\left(H_{uj}^{(k)}(\nu)\right)^{a_{H,uj}^{(k)}(\nu)-1}\label{eq:app-prior-distributions}
\end{eqnarray}

Furthermore, using the parameter independence, the prior distribution
for $\mathbf{\Theta}$ is,

\begin{equation}
p(\mathbf{\Theta})=p(\pi)p(\pmb{\rho})p(\mathbb{\mathbf{P}})p(\mathbb{\mathbf{H}})\label{eq:net-prior}
\end{equation}

Plugging Eq.~\ref{eq:net-prior} and Eq.~\ref{eq:prob y given ck theta}
into Eq.~\ref{eq:Q function def for semi markov}, and using the
hyperparameters mentioned in Sec.~\ref{subsec:EM-Algorithm}, we
get,

\begin{eqnarray}
Q(\mathbf{\Theta}|\mathbf{\Theta}^{(p)}) & = & \mathbb{E}_{\Theta^{(p)}}\left[\log(p({\bf Y}|\mathbf{\Theta})p(\mathbf{\Theta})\right]\nonumber \\
 & = & \sum_{n=1}^{N}\sum_{k\in\mathcal{K}}\Omega_{nk}(\mathbf{\Theta}^{(p)})\log\left[\pi^{(k)}p_{\mathbf{\Theta}}(\mathbf{y}^{(n)}|z^{(n)}=k)\right]+\log p(\mathbf{\Theta})\nonumber \\
 & = & \sum_{n=1}^{N}\sum_{k\in\mathcal{K}}\Omega_{nk}(\mathbf{\Theta}^{(p)})\log\left[\pi^{(k)}\rho_{u_{1}}^{(k)}\prod_{l=1}^{L^{(n)}}\left\{ P_{u_{l},u_{l+1}}^{(k)}\cdot H_{u_{l},u_{l+1}}^{(k)}(\nu_{l})\right\} \right]+\log p(\pi)p(\pmb{\rho})p(\mathbb{\mathbf{P}})p(\mathbb{\mathbf{H}})\nonumber \\
 & = & \sum_{n=1}^{N}\sum_{k\in\mathcal{K}}\log\left[\left(\pi^{(k)}\right)^{\Omega_{nk}(\mathbf{\Theta}^{(p)})}\left(\rho_{u_{1}}^{(k)}\right)^{\Omega_{nk}(\mathbf{\Theta}^{(p)})}\cdot\right.\nonumber \\
 &  & \left.\prod_{l=1}^{L^{(n)}}\left\{ \left(P_{u_{l},u_{l+1}}^{(k)}\right)^{\Omega_{nk}(\mathbf{\Theta}^{(p)})}\cdot\left(H_{u_{l},u_{l+1}}^{(k)}(\nu_{l})\right)^{\Omega_{nk}(\mathbf{\Theta}^{(p)})}\right\} \right]+\log p(\pi)p(\pmb{\rho})p(\mathbb{\mathbf{P}})p(\mathbb{\mathbf{H}})\nonumber \\
 & \propto & \log\left[\prod_{k\in\mathcal{K}}\left(\pi^{(k)}\right)^{\left(\sum_{n=1}^{N}\Omega_{nk}(\mathbf{\Theta}^{(p)})+a_{\pi}^{(k)}-1\right)}\right]+\nonumber \\
 &  & \sum_{k\in\mathcal{K}}\log\left[\prod_{u\in\mathcal{U}}\left(\rho_{u}^{(k)}\right)^{\left(\sum_{n=1}^{N}\Omega_{nk}(\mathbf{\Theta}^{(p)})\kappa(u_{1},u)+a_{\rho,u}^{(k)}-1\right)}\right]+\nonumber \\
 &  & \sum_{k\in\mathcal{K}}\sum_{u\in\mathcal{U}}\log\left[\prod_{j\in\mathcal{U}}\left(P_{uj}^{(k)}\right)^{\left(\sum_{n=1}^{N}\Omega_{nk}(\mathbf{\Theta}^{(p)})\bar{\kappa}_{uj}(\mathbf{y}^{(n)})+a_{P,uj}^{(k)}-1\right)}\right]+\nonumber \\
 &  & \sum_{k\in\mathcal{K}}\sum_{u\in\mathcal{U}}\sum_{j\in\mathcal{U}}\log\left[\prod_{\nu\in\mathcal{T}}\left(H_{uj}^{(k)}(\nu)\right)^{\left(\sum_{n=1}^{N}\Omega_{nk}(\mathbf{\Theta}^{(p)})\tilde{\kappa}_{uj,\nu}(\mathbf{y}^{(n)})+a_{H,uj}^{(k)}(\nu)-1\right)}\right]\nonumber \\
 & \propto & \log\left[\pi^{(k)}\sim\text{Dirichlet}(\sum_{n=1}^{N}\Omega_{nk}(\mathbf{\Theta}^{(p)})+a_{\pi}^{(k)})\right]+\nonumber \\
 &  & \log\left[\rho_{u}^{(k)}\sim\text{Dirichlet}(\sum_{n=1}^{N}\Omega_{nk}(\mathbf{\Theta}^{(p)})\kappa(u_{1},u)+a_{\rho,u}^{(k)})\right]+\nonumber \\
 &  & \log\left[P_{uj}^{(k)}\sim\text{Dirichlet}(\sum_{n=1}^{N}\Omega_{nk}(\mathbf{\Theta}^{(p)})\bar{\kappa}_{uj}(\mathbf{y}^{(n)})+a_{P,uj}^{(k)})\right]+\nonumber \\
 &  & \log\left[H_{uj}^{(k)}(\nu)\sim\text{Dirichlet}(\sum_{n=1}^{N}\Omega_{nk}(\mathbf{\Theta}^{(p)})\tilde{\kappa}_{uj,\nu}(\mathbf{y}^{(n)})+a_{H,uj}^{(k)}(\nu))\right]\label{eq:posterior-derivation}
\end{eqnarray}

where, $\kappa(x,y)$ is an indicator function equal to 1 if $x=y$,
$\bar{\kappa}_{uj}(\mathbf{y}^{(n)})$ is the count function equal
to the number of times transition was made from state $u$ to $j$
in trajectory $\mathbf{y}^{(n)}$, and $\tilde{\kappa}_{uj,\nu}(\mathbf{y}^{(n)})$
is the count function equal to the number of times transition was
made from state $u$ to $j$, in trajectory $\mathbf{y}^{(n)}$, when
length of stay at state $u$ was $\nu$ time units.

As shown in Eq.~\ref{eq:posterior-derivation}, the posteriors of
the model parameters are Dirichlet distributions with updated hyperparameters.
The posterior of any Dirichlet variable, $x_{1,}\ldots,x_{m}\sim\text{Dirichlet}(a_{1},\ldots,a_{m})$
is maximized at $E[x_{i}]=\cfrac{a_{i}}{\sum_{i'=1}^{m}a_{i'}},\forall i$.
Thus, the parameter estimates to maximize Eq.~\ref{eq:Q function def for semi markov}
are,

\begin{eqnarray*}
\pi^{(k)(p+1)} & = & \frac{\sum_{n=1}^{N}\Omega_{nk}(\mathbf{\Theta}^{(p)})+a_{\pi}^{(k)}}{\sum_{k'\in\mathcal{K}}\left[\sum_{n=1}^{N}\Omega_{nk'}(\mathbf{\Theta}^{(p)})+a_{\pi}^{(k')}\right]},\forall k\in\mathcal{K}.\\
\rho_{u}^{(k)(p+1)} & = & \frac{\sum_{n=1}^{N}\Omega_{nk}(\mathbf{\Theta}^{(p)})\kappa(u_{1},u)+a_{\rho,u}^{(k)}}{\sum_{u'\in\mathcal{U}}\left[\sum_{n=1}^{N}\Omega_{nk}(\mathbf{\Theta}^{(p)})\kappa(u_{1},u')+a_{\rho,u'}^{(k)}\right]},\forall u\in\mathcal{U},k\in\mathcal{K}\\
P_{uj}^{(k)(p+1)} & = & \frac{\sum_{n=1}^{N}\Omega_{nk}(\mathbf{\Theta}^{(p)})\bar{\kappa}_{uj}(\mathbf{y}^{(n)})+a_{P,uj}^{(k)}}{\sum_{j'\in\mathcal{U}}\left[\sum_{n=1}^{N}\Omega_{nk}(\mathbf{\Theta}^{(p)})\bar{\kappa}_{uj'}(\mathbf{y}^{(n)})+a_{P,uj'}^{(k)}\right]},\forall u,j\in\mathcal{U},k\in\mathcal{K}\\
H_{uj}^{(k)}(\nu)^{(p+1)} & = & \frac{\sum_{n=1}^{N}\Omega_{nk}(\mathbf{\Theta}^{(p)})\tilde{\kappa}_{uj,\nu}(\mathbf{y}^{(n)})+a_{H,uj}^{(k)}(\nu)}{\sum_{\nu'\in\mathcal{T}}\left[\sum_{n=1}^{N}\Omega_{nk}(\mathbf{\Theta}^{(p)})\tilde{\kappa}_{uj,\nu'}(\mathbf{y}^{(n)})+a_{H,uj}^{(k)}(\nu')\right]},\forall u,j\in\mathcal{U},\nu\in\mathcal{T},k\in\mathcal{K}
\end{eqnarray*}

\subsection*{Appendix B: Elective Scheduling Optimization MIP Formulation}

\label{app:elecopt} In this appendix we present, an optimization
model from the literature (Helm and Van Oyen (2015)). that is used
to demonstrate the importance of a rigorous patient trajectory estimation
procedure. We designed our estimation approach to integrate with optimization
and what-if scenarios, with this particular optimization being used
as a proof of concept that (1) our method integrates well with current
optimization approaches, and (2) our method significantly improves
the outcome of the optimization when compared with traditional approaches
proposed for use with these types of models. We begin by describing
the model parameters and then present the optimization model with
brief description of the objective and constraints. For a more detailed
description of the optimization approach we refer the readers to Helm
and Van Oyen (2015).

\begin{singlespace}
\begin{tabular}{ll}
\multicolumn{2}{l}{\textbf{Sets}}\tabularnewline
$\mathcal{K}$  & set of all patient types\tabularnewline
$\mathcal{U}$  & set of hospital wards\tabularnewline
 & \tabularnewline
\multicolumn{2}{l}{\textbf{\medskip{}
 Hospital parameters}}\tabularnewline
$\mathbf{{\bf \text{\textgreek{z}}}}$  & vector of ward capacities\tabularnewline
${\bf {\eta}}$  & vector of total cancellations attributed for each ward\tabularnewline
$b$  & limit on the average number of blockages per week\tabularnewline
$\mathbf{o}$  & vector of limit on the average number of off-unit patients allowed
for each ward\tabularnewline
$\mu_{d}^{(k)}$  & current elective admission volume of type $k$ patients on day $d$\tabularnewline
$\bar{\mu}_{d}^{(k)}$  & maximum number of elective admissions of type $k$ allowed on day
$d$\tabularnewline
$\mathbf{R}$  & reward vector where $R_{k}$ is the reward for admitting patient of
type $k$\tabularnewline
 & \tabularnewline
\multicolumn{2}{l}{\textbf{\medskip{}
 Patient trajectory and census distributions}}\tabularnewline
$\gamma_{u}^{(k)}(d_{1})$  & probability that an elective patient of type $k$ requires a bed in
ward $u$, $d_{1}$ days after \tabularnewline
 & admission (trajectory distribution)\tabularnewline
$p_{u,d}(n)$  & probability that there are $n$ emergency patients demanding a bed
in ward $u$ on day $d$\tabularnewline
$\bar{p}_{d}(n)$  & probability that there are $n$ emergency patients demanding a bed
in the hospital on day $d$\tabularnewline
 & \tabularnewline
\multicolumn{2}{l}{\textbf{\medskip{}
 Decision Variables}}\tabularnewline
$\Psi_{d}^{(k)}$  & number of type $k\in\mathcal{K}$ patients scheduled on day $d$\tabularnewline
$\delta_{d,n}$  & number of blockages if there are $n$ emergency patients in the hospital
on day $d$\tabularnewline
$\text{\textgreek{'o}}_{d,n}^{u}$  & number of ward $u$ off-unit patients on day $d$ if there are $n$
emergency patients in ward $u$\tabularnewline
\end{tabular}
\end{singlespace}

\medskip{}

The patient trajectory and census distribution parameters are computed
offline as explained earlier in this section. Since the PALM model
for emergency patient bed demand is exogenous to the decision variable,
this too is calculated off-line, with the results captured as $p_{u,d}(n)$
and $\bar{p}_{d}(n)$. We consider a weekly planning horizon that
repeats itself every week, generating a cyclostationary system that
varies by day of week. The objective is to maximize the throughput
of the sum of elective patient admissions (over the planning horizon)
of each type weighted by a \textquotedbl{}reward\textquotedbl{} vector
$\mathbf{R}$ ($\mathbf{1}$ denotes a column vector of all ones).
The reward vector gives flexibility to allow the model to treat one
patient type differently from another, for example, the model can
prioritize one patient type over another with respect to patient criticality,
projected revenue generated by the admission, or other strategic priority.
The formulation is as follows: 
\begin{flalign}
\max_{\Theta,\delta,\hat{\delta}}\mathbf{R}\cdot\Psi\cdot\mathbf{1}\label{eq:MaxElec_Obj}
\end{flalign}
$s.t.$

\begin{flushleft}
\begin{align}
\delta_{d_{1},n} & \geq n-\sum_{u\in\mathcal{U}}(\text{\textgreek{z}}{}_{u}-\sum_{d_{2}=1}^{7}\sum_{k\in\mathcal{K}}\Psi_{d_{2}}^{(k)}\cdot\sum_{n'=0}^{\infty}\gamma_{u}^{(k)}(7n'+d_{1}-d_{2})),\hspace{1em}\hspace{1em}\hspace{1em}\hspace{1em}\label{eq:MaxElec_Cons1_Exp_blockages}\\
 & \hspace{1em}\hspace{1em}\hspace{1em}\hspace{1em}\hspace{1em}\hspace{1em}\hspace{1em}\hspace{1em}\hspace{1em}\hspace{1em}d_{1}=1,\ldots,7;\:n=1,2,\ldots\nonumber \\
\sum_{d=1}^{7}\sum_{n=0}^{\infty}\bar{p}_{d}(n)\delta_{d,n} & \leq b\label{eq:MaxElec_Cons2_WklyBlck-1}\\
\delta_{d,n+1} & \geq\delta_{d,n}\hspace{1em}\hspace{1em}\hspace{1em}\hspace{1em}\hspace{1em}\hspace{1em}\hspace{1em}d=1,\ldots,7;\:n=1,2,\ldots\label{eq:MaxElec_Cons3_Cuts-1}\\
\text{\textgreek{'o}}_{d_{1},n}^{u} & \geq n+\sum_{d_{2}=1}^{7}\sum_{k\in\mathcal{K}}\Psi_{d_{2}}^{(k)}\cdot\sum_{n'=0}^{\infty}\gamma_{u}^{(k)}(7n'+d_{1}-d_{2})-\text{\textgreek{z}}{}_{u}-\eta_{u}\sum_{d=0}^{7}\sum_{n'=0}^{\infty}\delta_{d,n'}\cdot\bar{p}_{d}(n')\label{eq:MaxElec_Cons4-1}\\
 & \hspace{1em}\hspace{1em}\hspace{1em}\hspace{1em}\hspace{1em}\hspace{1em}\hspace{1em}\hspace{1em}\hspace{1em}\hspace{1em}\forall u\in\mathcal{U};\:d_{1}=1,\ldots,7;\:n=1,2,\ldots\nonumber \\
\sum_{n=0}^{\infty}p_{u,d}(n)\text{\textgreek{'o}}_{d,n}^{u} & \leq\mathbf{o}_{u}\hspace{1em}\hspace{1em}\hspace{1em}\hspace{1em}\hspace{1em}\hspace{1em}\hspace{1em}\forall u\in\mathcal{U};\:d=1,\ldots,7\label{eq:MaxElec_Cons5-1}\\
\text{\textgreek{'o}}_{d,n+1}^{u} & \geq\text{\textgreek{'o}}_{d,n}^{u}\hspace{1em}\hspace{1em}\hspace{1em}\hspace{1em}\hspace{1em}\hspace{1em}\hspace{1em}d=1,\ldots,7;\:n=1,2,\ldots\label{eq:MaxElec_Cons6-1}\\
\sum_{d=1}^{7}\Psi_{d}^{(k)} & \geq\sum_{d=1}^{7}\mu_{d}^{(k)}\hspace{1em}\hspace{1em}\hspace{1em}\hspace{1em}\forall k\in\mathcal{K}\label{eq:MaxElec_Cons7-1}\\
\Psi_{d}^{(k)} & \leq\bar{\mu}_{d}^{(k)}\hspace{1em}\hspace{1em}\hspace{1em}\hspace{1em}\hspace{1em}\hspace{1em}\forall k\in\mathcal{K};\:d=1,\ldots,7\label{eq:MaxElec_Cons8-1}
\end{align}
\par\end{flushleft}

\begin{align*}
\Psi_{d}^{(k)},\delta_{d,n},\text{\textgreek{'o}}_{d,n}^{u} & \in\mathbb{Z}^{+}
\end{align*}

\medskip{}

The constraints of this model are primarily for constraining the blockages
faced by the patients, limiting off-ward placement, and respecting
the hospital resource limits. Since the purpose of this work is to
demonstrate how CM can be improved by developing methods that integrate
with optimization, and not to provide new optimization methods, we
briefly describe the optimization presented here. Greater detail regarding
this approach can be found in Helm and Van Oyen (2015). Constraints
\ref{eq:MaxElec_Cons1_Exp_blockages} calculate the number of blocked
patients at the hospital level if $n$ emergency patients are in the
hospital on day $d_{1}$. This sets the helper variable, $\delta_{d,n}$
which is subsequently used to calculate expected blockages according
to the distribution on the emergency patient bed demand stochastic
process in the left hand side (LHS) of Constraints \ref{eq:MaxElec_Cons2_WklyBlck-1}
by multiplying the indicator of whether the $n^{th}$ patient would
be blocked by the probability of seeing $n$ emergency patients in
the hospital. The right hand side constrains the expected blocked
patients to be less than some target level, $b$, which can be chosen
by management. Constraint \ref{eq:MaxElec_Cons3_Cuts-1} is a cut
that is added to the formulation that significantly improves model
solution speed.

Similar to the constraints (Eq.~\ref{eq:MaxElec_Cons1_Exp_blockages}-\ref{eq:MaxElec_Cons3_Cuts-1})
for blockages, we have constraints in Eq.~\ref{eq:MaxElec_Cons4-1}-\ref{eq:MaxElec_Cons6-1}
for approximating and limiting expected off-unit census. An additional
term in Eq.~\ref{eq:MaxElec_Cons4-1}, $\eta_{u}\sum_{d=0}^{7}\sum_{n'=0}^{\infty}\delta_{d,n'}\cdot\bar{p}_{d}(n')$,
subtracted from the otherwise expected number of off-unit census gives
patients who were blocked and not able to be admitted to the hospital
in the first place.

Constraints in Eq.~\ref{eq:MaxElec_Cons7-1} ensures that the proper
mix of patients is respected. Specifically, it ensures that each patient
type has at least as many admissions each week as they did prior to
optimization. Constraints \ref{eq:MaxElec_Cons8-1} ensure that the
model respects the hospital resource capacity for a day. For example,
hospitals frequently avoid admitting elective patients on Sundays,
which could be achieved by setting $\bar{\mu}_{Sunday}^{(k)}=0$.

\subsection*{Appendix C: Assigning attributes to patients in simulation study}

\label{subsec:Appendix-C-Simulation-attribute-assignment}

Here we elaborate on synthesis patient attributes for simulation study
in Sec.~\ref{sec:Validation-using-Simulation}. For brevity, we show
it for $K=4$. In the data generation step for this problem, after
patient trajectories were generated from four semi-Markov processes,
three attributes, viz. age, gender and diagnosis (with three diagnoses
being D1, D2, D3), were assigned to the patients such that any attribute
triplet has the possibility of being in any cluster; e.g. a 30 year
old female with diagnosis D1 could potentially be from any of the
four clusters. This resembles real-world challenges involved in patient
trajectory estimation by simulating the fact that two patients with
the same attributes may have different trajectories; i.e. the attributes
are not adequately capturing patient heterogeneity. In practice, patient
attributes are capable of capturing some of the patient heterogeneity
so we ensure that clusters contain patients whose attributes are mostly
similar by adhering to a near-Pareto principle (see the three attribute
generating tables in Table~\ref{tab:Generating-distributions-for-patient-attributes}).
That is, clusters are composed mostly of similar patient attributes
with a mix of patients who have different attributes. This distribution
of attributes is designed to be fair to the traditional approach and
capture the reality that attributes do have differentiating power,
but cannot completely specify a patients likely trajectory. 

In Sec.~\ref{subsec:Appendix-C-Simulation-attribute-assignment},
we assign physical attributes to patient for our simulation study.
We perform a conservative assignment, in favor of traditional patient
clustering method, by giving higher chance of patients within a \textit{true}
cluster having similar attributes. Table~\ref{tab:Generating-distributions-for-patient-attributes}
below shows the generating distributions for the patient attributes
within each cluster. As shown in the table, age is taken from a normal
distribution with different means (Table~\ref{tab:Normal-Distribution-for-age}),
sex and diagnosis (Table~\ref{tab:Uniform-distribution-for-sex}-\ref{tab:Uniform-distribution-for-diagnosis})
are taken according to a Bernoulli random variable with different
success probabilities. The distribution parameters are chosen such
that there is high attribute similarity (dissimilarity) between patients
within (between) clusters.

\begin{table}
\begin{centering}
\begin{minipage}[b][1\totalheight][t]{0.2\columnwidth}%
\begin{center}
\subfloat[Normal Distribution for patient age \label{tab:Normal-Distribution-for-age}]{\begin{centering}
\begin{tabular}{c|c}
\textbf{\footnotesize{}Cluster } & \textbf{\footnotesize{}Age}\tabularnewline
\hline 
{\footnotesize{}1 } & {\footnotesize{}$~N(20,3)$}\tabularnewline
{\footnotesize{}2 } & {\footnotesize{}$~N(30,3)$}\tabularnewline
{\footnotesize{}3 } & {\footnotesize{}$~N(40,3)$}\tabularnewline
{\footnotesize{}4 } & {\footnotesize{}$~N(50,3)$}\tabularnewline
\hline 
\end{tabular}
\par\end{centering}
}
\par\end{center}%
\end{minipage}\hspace{1cm}%
\begin{minipage}[b][1\totalheight][t]{0.3\columnwidth}%
\begin{center}
\subfloat[Uniform distribution for patient sex within clusters\label{tab:Uniform-distribution-for-sex}]{\begin{centering}
\begin{tabular}{c|c|c}
 & \multicolumn{2}{c}{{\footnotesize{}Sex}}\tabularnewline
\hline 
\textbf{\footnotesize{}Cluster } & \textbf{\footnotesize{}M } & \textbf{\footnotesize{}F}\tabularnewline
\hline 
{\footnotesize{}1 } & {\footnotesize{}80\% } & {\footnotesize{}20\%}\tabularnewline
{\footnotesize{}2 } & {\footnotesize{}20\% } & {\footnotesize{}80\%}\tabularnewline
{\footnotesize{}3 } & {\footnotesize{}70\% } & {\footnotesize{}30\%}\tabularnewline
{\footnotesize{}4 } & {\footnotesize{}30\% } & {\footnotesize{}70\%}\tabularnewline
\hline 
\end{tabular}
\par\end{centering}
}
\par\end{center}%
\end{minipage}\hspace{1cm}%
\begin{minipage}[b][1\totalheight][t]{0.3\columnwidth}%
\begin{center}
\subfloat[Uniform distribution for patient diagnosis\label{tab:Uniform-distribution-for-diagnosis}]{\begin{centering}
\begin{tabular}{c|ccc}
 & \multicolumn{3}{c}{{\footnotesize{}Diagnosis}}\tabularnewline
\hline 
\textbf{\footnotesize{}Cluster } & \textbf{\footnotesize{}D1 } & \textbf{\footnotesize{}D2 } & \textbf{\footnotesize{}D3}\tabularnewline
\hline 
{\footnotesize{}1 } & {\footnotesize{}70\% } & {\footnotesize{}20\% } & {\footnotesize{}10\%}\tabularnewline
{\footnotesize{}2 } & {\footnotesize{}20\% } & {\footnotesize{}70\% } & {\footnotesize{}10\%}\tabularnewline
{\footnotesize{}3 } & {\footnotesize{}10\% } & {\footnotesize{}20\% } & {\footnotesize{}70\%}\tabularnewline
{\footnotesize{}4 } & {\footnotesize{}80\% } & {\footnotesize{}10\% } & {\footnotesize{}10\%}\tabularnewline
\hline 
\end{tabular}
\par\end{centering}
}
\par\end{center}%
\end{minipage}
\par\end{centering}
\caption{Generating distributions for patient attributes within \textit{true
}clusters\label{tab:Generating-distributions-for-patient-attributes}}
\end{table}

\subsection*{Appendix D: Empirical Estimation of Patient Trajectories}

\label{app:empirical} Once the clusters have been formed using k-means
clustering, the trajectory distribution is computed for each cluster
independently by normalizing the frequency of transitions of patients
between wards as follows:

\vspace{-0.2in}
 
\begin{enumerate}
\item $\rho_{u}^{(k)}=\frac{\sum_{n=1}^{N}\kappa(\mathbf{y}_{1}^{(n)},u)}{\sum_{u'\in\mathcal{U}}\left[\sum_{n=1}^{N}\kappa(\mathbf{y}_{1}^{(n)},u')\right]}$for
$k\in\mathcal{K}$ and $u\in\mathcal{U}$. 
\item $P_{uj}^{(k)}=\frac{\sum_{n=1}^{N}\bar{\kappa}_{uj}(\mathbf{y}^{(n)})}{\sum_{j'\in\mathcal{U}}\left[\sum_{n=1}^{N}\bar{\kappa}_{uj'}(\mathbf{y}^{(n)})\right]}$
for $k\in\mathcal{K};\:u\in\mathcal{U}$ and $j\in\mathcal{U}$. 
\item $H_{uj}^{(k)}(\nu)=\frac{\sum_{n=1}^{N}\tilde{\kappa}_{uj,\nu}(\mathbf{y}^{(n)})}{\sum_{\nu'\in\mathcal{T}}\left[\sum_{n=1}^{N}\tilde{\kappa}_{uj,\nu'}(\mathbf{y}^{(n)})\right]}$
for $k\in\mathcal{K},u\in\mathcal{U}$, $j\in\mathcal{U}$ and $\nu\in\mathcal{T}$. 
\end{enumerate}
\vspace{-0.2in}

\end{document}